\documentclass[journal,transmag]{IEEEtran}

\usepackage[usenames,dvipsnames,svgnames,table]{xcolor}
\usepackage{graphicx}
\usepackage{subfig}
\usepackage{url}
\usepackage{microtype} % better typesetting
\usepackage[english]{babel} % typographical rules
\usepackage{amsmath}
\usepackage[ruled,noend,slide,lined]{algorithm2e}
\usepackage{setspace} %to change space in algorithm

\usepackage{hyperref}
\hypersetup{
	pdffitwindow=false,
	hidelinks,
	bookmarksopen=true
}
\usepackage{hypcap}

\urldef{\mailsa}\path|{bernd.meijerink, m.baratchi, geert.heijenk}@utwente.nl|

\begin{document}
%%%%%%%%%%%%%%%%%%%%
\title{A Distributed Routing Algorithm for Internet-wide Geocast}

\author{
\IEEEauthorblockN{ 	Bernd Meijerink\IEEEauthorrefmark{1},
					Mitra Baratchi\IEEEauthorrefmark{2}, and
					Geert Heijenk\IEEEauthorrefmark{1} }
\IEEEauthorblockA{ 	\IEEEauthorrefmark{1}University of Twente, Enschede, The Netherlands }
\IEEEauthorblockA{ 	\IEEEauthorrefmark{2}Leiden Institute of Advanced Computer Science (LIACS), Leiden University, Leiden, The Netherlands }
}

% abstract and (if needed) index terms
\IEEEtitleabstractindextext{%
	\begin{abstract}
		Geocast is the concept of sending data packets to nodes in a specified geographical area instead of nodes with a specific address. To route geocast messages to their destination we need a geographic routing algorithm that can route packets efficiently to the devices inside the destination area. 
		Our goal is to design an algorithm that can deliver shortest path tree like forwarding while relying purely on distributed data without central knowledge.
		In this paper, we present two algorithms for geographic routing. One based purely on distance vector data, and one more complicated algorithm based on path data. In our evaluation, we show that our purely distance vector based algorithm can come close to shortest path tree performance when a small number of routers are present in the destination area. We also show that our path based algorithm can come close to the performance of a shortest path tree in almost all geocast situations.
	\end{abstract}
}

\maketitle
\IEEEdisplaynontitleabstractindextext

%%%%%%%%%%%%%%%%%%%%
%keywords: routing, geocast, geographical routing

\section{Introduction}
\label{sec:introduction}

Due to the increasing use of connected, possibly autonomous vehicles and `smart' devices there is an increasing need for Internet-wide geographically scoped communications \cite{karagiannis2013internet}. This would allow devices in a specific area to be addressed without the need of keeping track of all IP addresses or administration concerning multicast groups. This can be achieved through geocasting.

The main idea behind geocast is to route packets based on a geographic destination area instead of a fixed address or multicast group \cite{conf/mobicom/NavasI97}. %Routers forward packets based on the destination area
This could allow the transmission of data towards devices in a region without complex bookkeeping of device locations. 

\begin{figure}[b]
	\centering
	\includegraphics[width=0.9\linewidth, clip=true, trim=2mm 38mm 120mm 145mm]{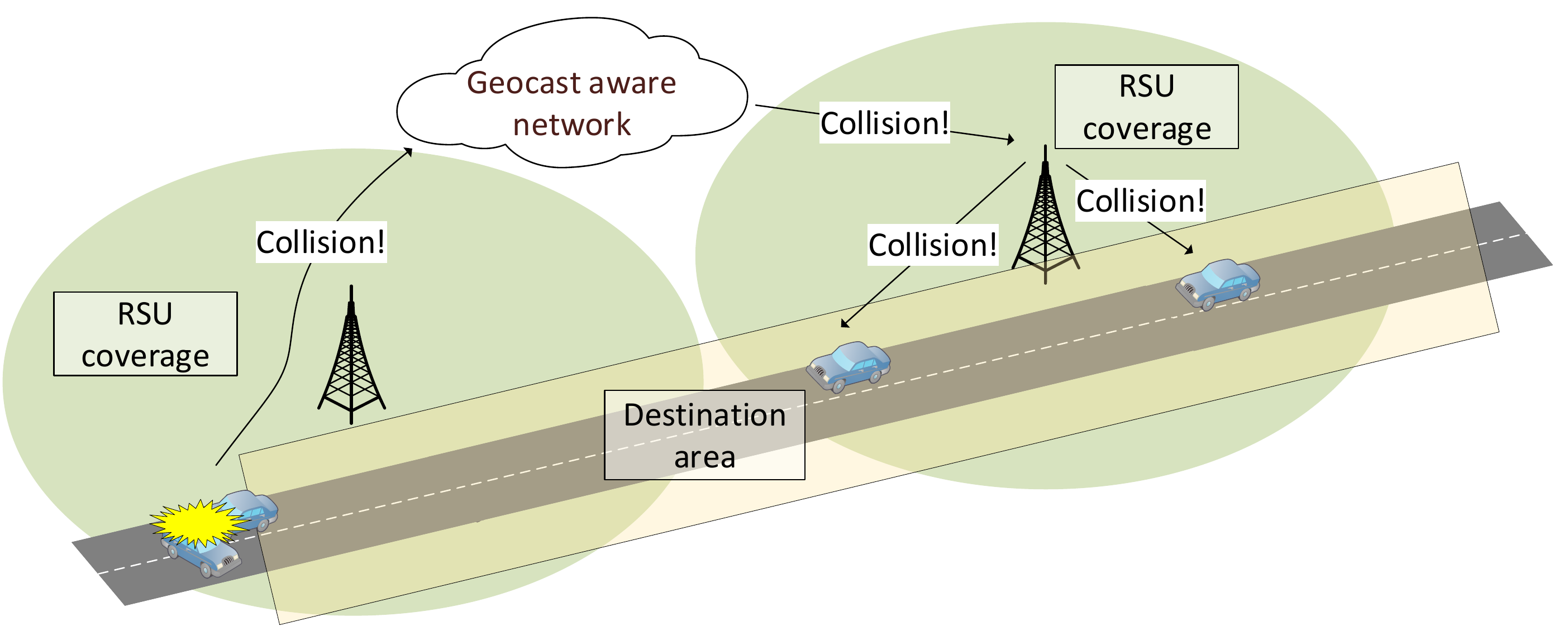}
	\caption{Geocast traffic accident example}
	\label{fig:car_example_rsu}
\end{figure}

We mostly focus on the possible applications of geocast in the vehicular network domain. Possible applications would be location dependent weather updates, traffic alerts and information to assist autonomous driving.
In the vehicular networking scenario a functioning geocast solution would at least require location aware-base stations, or Road Side Units (RSUs) that are aware of the area they serve. Location aware-systems are important for many safety applications related to vehicular networks \cite{CunhaVillasBoukercheEtAl2016}.
One example of a vehicular networking geocast scenario would be notifying cars on a specific highway of an accident or traffic jam, using geocast to only send the message to that street. An example of such a situation is depicted in Figure \ref{fig:car_example_rsu} were a RSU forwards a notification of a traffic accident to multiple vehicles on the road of the accident.

Currently available implementations of geocast are mostly application layer based, an example being extended DNS\cite{conf/vnc/FiorezeH11}. There are two main downsides to such an approach. They have high overhead due to lookup operations and are less resilient to change. We propose an alternative approach to the problem: Implementing geocast on the network layer. A network layer implementation would allow us to use information already available due to unicast routing. The system would also be more resilient due to not relying on availability of certain servers and embedding geocast in the network itself will allow it to route around problems in the network. It would also enable such a system to possibly scale to the entire Internet. Enabling Internet-wide geocast could potentially allow fine grained geographically scoped message transmission for everyone. The main benefit would be that sending hosts on the network do not require any sort of geographical information, they can just send a geocast packet to the router serving them. Possible use cases of network layer geocast range from localized weather reports without clients reporting their location to safety information transmitted to vehicles.

To provide an efficient geocasting solution, the underlying routing protocol will need to take geographic information into account. Traditional routing methods such as unicast or multicast routing have drawbacks in the geocast scenario.

Unicast routing has the obvious drawback of sending one packet per destination. This would lead to communications overhead with a large number of devices in the destination area. On the other hand, the per-packet processing overhead is minimal as unicast routing is well understood and optimized.

Multicast routing seems like a better fit as it already supports one-to-many communications. The main drawback for geographically scoped communication is that most multicast routing solutions depend on subscription messages. In a geocast solution routers will need to report their coverage area and evaluate which routers cover the destination area of a packet. another mayor drawback would the be requirement of predefined destination areas, as it would have to be known which multicast group covers which area.

We set out to design a distributed algorithm as the communication overhead needed for a centralized approach would lead to scalability problems for the system. Routers should not depend on some central authority or need full network knowledge. This last requirement prevents us from constructing a least cost (Steiner) tree, as this would require full network knowledge. We will cover this last point in greater detail in Section \ref{sec:prev_work:sp}.

For efficient geographic routing we need a routing algorithm in which geographical areas are central to packet routing. A geographic routing algorithm will need to efficiently route packets that have a geographic destination to all routers that (partially) cover that area. We specifically refer to coverage instead `being in the area', as the important thing is that devices connected to the router are in the destination area.
The most important aspect is the ability to route a packet to multiple destinations using the lowest number of hops possible, without sending duplicate packets over the same link.

%TODO dit naar background moven?
We use two area definitions in our geocast system: Coverage area and destination area.
\textit{Coverage Area} defines the geographic area that is covered by a router, devices in this area can be reached through this router. Coverage areas of routers may overlap or even be identical, for example multiple providers servicing the same area.
\textit{Destination area} refers to the geographic area to which a packet is sent. This area does not need to be identical to the coverage area of a router, instead routers should calculate if the destination overlaps with their coverage area.

The main research question we answer in this paper is: How can we efficiently route geocast packets within a network.
The main contribution of our work is threefold:
\begin{itemize}
	\item We design an efficient geographical routing algorithm based on path information,
	\item We design a geographical routing algorithm using purely distance vector based information,
	\item We validate and evaluate the proposed algorithms on a large set of real world network topologies.
\end{itemize}
Our second, less efficient algorithm is also less computationally intensive and can be used in cases where network capacity is less important than router resources.

The remainder of this paper is structured as follows: In the next section we will describe previous work in the area, including our own work on geographic addressing. Section \ref{sec:algorithms} will describe the algorithms we have designed to perform the geographic routing. We evaluate our algorithms in Section \ref{sec:evaluation}. Finally we draw our conclusions in Section \ref{sec:conclusion}.

\section{Previous Work}
\label{sec:prev_work}
In this section, we will describe previous work done on geocast and geographic routing. We start with describing related work in the wireless domain, moving towards work on geocast in a wired setting. We will conclude the section with an overview of our own work on geographic addressing and forwarding tree evaluation, in which we will explain our choice for a shortest path tree.

\subsection{Related work}
Geocast was initially introduced by Navas and Imielinski for wired networks \cite{conf/mobicom/NavasI97}. Their approach relied on special routers that know their location and forward packets based on the destination point, circle or polygon. Routers are connected hierarchically: A router that covers a certain area connects to `lower' routers that cover smaller areas within that coverage area. Routers can calculate the intersect of destination and their coverage using the GPS coordinates. Downsides of this approach are the hierarchical router requirement, the need for routers to perform area intersection calculations and the variable length of the addressing (points, circles, or polygons).

In later work from the same authors they studied improved routing cost \cite{navas2000reducing} by approximating the destination / coverage area intersection. They have also studied alternate approaches based on addressing predefined locations \cite{conf/mobide/ImielinskiG99}.

Most work on the topic of geocast has been done in the wireless ad-hoc network context, and especially the VANET case. Overviews of such routing protocols and underlying mechanism can be found in \cite{di2013group},\cite{SharefAlsaqourIsmail2014} and \cite{LiuWanWangEtAl2016}. In most of these protocols the location of forwarding nodes is tightly coupled with the destination of a packet, a next hop node will generally be in the direction of the target geocast area. The correlation between the position of the next hop node and the location of the destination area does not necessarily exist in a fixed wired network situation. Especially in situations were a network serves several access networks, there might be very little correlation between the forwarding routers and the actual destination area. On the other hand, the fixed wired environment is usually mostly static, this enables route distribution to be effective over long distances.

A well-known example of a geographic routing protocol for ad-hoc networks is GeoTORA \cite{KoVaidya2000}. When a node in the network needs to geocast a message it broadcasts a query with the request for the destination nodes. The destination nodes send a message back, allowing the original requesting node to know the forwarding hop towards the geocast area. The mobile nature of these ad-hoc networks makes this kind of signaling a necessity to reach any sort of efficiency. We do not have this problem in wired networks, allowing for the possibility of route distribution beforehand. 

Another example is Greedy Perimeter Stateless Routing \cite{Karp2000}. In this algorithm, traffic is routed to nodes that are located closer to the destination area than the transmitting node. This approach is seen often in geographic routing solutions for ad hoc networks. The downside for a wired environment is that the location of routers is not necessarily correlated with the direction a packet needs to travel to reach its destination.

There are off-course algorithms for multicast routing such as Protocol Independent Multicast (PIM). These could be used in some capacity for geocast routing but they do have some drawbacks. PIM Dense Mode (PIM-DM) relies on an initial flooding stage where routers that are not subscribers send a prune message back to their forwarding neighbor \cite{FarinacciLiHanksEtAl2005}. We would ideally like to not have this behaviour in our geocast system as we believe the number of destination areas that might be addressed in a short time could be very large. Alternatively, PIM Sparse Mode (PIM-SM) relies on an initial Rendezvous Router that routes packets before a shortest path tree is established \cite{FennerHandleyKouvelasEtAl2006}. Due to the large number of varying geocast destinations and the overhead caused by the Rendezvous Router we believe this approach would not be feasible.

Another approach to geocast is to use DNS to resolve geographical areas to a IP addresses by extending the DNS \cite{conf/vnc/FiorezeH11}. When the eDNS server is queried for a certain area, it returns the IP addresses of all entries in that region. The eDNS was designed for VANET scenarios, so it would only have to return a list of RSUs in the target area. Scaling the system to track the movements of all vehicles to also allow geocasting from multiple networks at the same time was later found to be somewhat feasible \cite{MoscoviterGholibeigiMeijerinkEtAl2016}. For a truly Internet-wide deployment such a system would need to scale significantly. DNS like delegation would also be complicated if updates are to be distributed through the network in a relatively short time.

\subsection{Geographic addressing}
\label{sec:prev_work:geo_addressing}
\begin{figure}[b]
	\centering
	\includegraphics[width=1.0\linewidth, clip=true, trim=6mm 6mm 6mm 6mm]{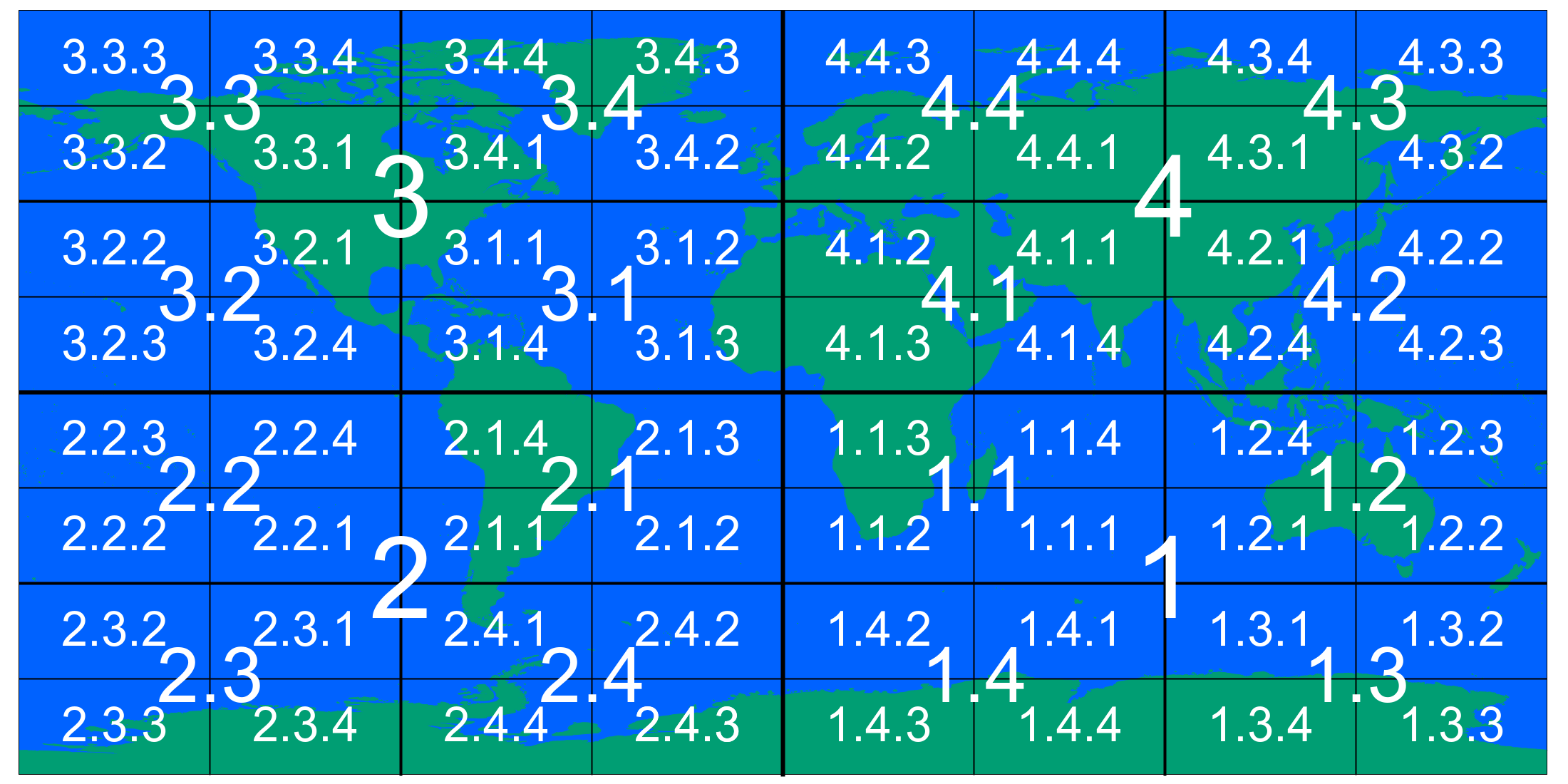}
	\caption{Geographic addressing up to level 3}
	\label{fig:levels}
\end{figure}

Our view is that a geographic addressing scheme is needed to allow efficient geographic routing in the Internet and thus as an extension geocast itself.
In our previous work, we described an addressing scheme for geocast in the Internet \cite{meijerink2016efficient}.
Our addressing scheme transforms a location bounding box defined by its maximum and minimum latitude and longitude into an IPv6 address. This is achieved by dividing the area of the world into 4 rectangles, and in turn subdividing these rectangles. We number the rectangles in such a way that neighboring rectangles with different `parents' share the same number. An example of this method mapped to a world map with a depth of 3 levels can be seen in Figure \ref{fig:levels}. Using this hierarchical structure we can address increasingly smaller blocks for more precision down to areas with a size of 7 by 3.5 cm.

This addressing scheme effectively decouples the actual geographic location and the address itself. Routers or other forwarding systems do not need to have any knowledge of their geographic location, only of the underlying addressing system. Our addressing system allows the matching of a destination area and a coverage area in a manner similar to prefix matching. This is achieved by encoding each level in a block of 4 bits. We can perform a bitwise AND operation on the destination and coverage area to find if there is overlap. Overlap is found if there is at least one bit shared in each 4 bit group, up until the length of the shortest address (which corresponds to the largest area).

As an example address we will take the area of the city of Enschede, bounded by 52.24 degrees north, 6.94 degrees east, 52.19 degrees south and 6.84 degrees west. This area is completely covered by the 12 level address 4.4.2.3.2.1.1.2.4.[2,3].[1,2].4 (with [2,3] meaning we address both rectangle 2 and 3), which has the binary representation `0001 0001 0100 0010 0100 1000 1000 0100 0001 0110 1100 0001'. In turn this gives us the following IPv6 hexadecimal representation: `1142:4884:16c1::'. For an actual deployment we would need to add a prefix to distinguish the geocast addresses from unicast or multicast addresses.

The addressed area does have to be symmetrical, this might cause the addressed area to be greater than the actual destination area. We have, however, shown that our addressing scheme will allow a packet to efficiently get close to its destination with minimal processing overhead \cite{meijerink2016efficient}. Once the packet reaches the final router in its destination, a more accurate distribution system might have to take over (for example, one specific for vehicular networks).

\subsection{Why shortest path instead of a Steiner tree}
\label{sec:prev_work:sp}
In an ideal world we would always transmit packets using the least cost tree (Steiner tree) from source to destinations. By definition this is the best routing tree that can be established based on chosen metrics such as cost or delay. There are however several drawbacks to such an approach that have real world implications. Some of these drawbacks are: The requirement of full network knowledge and high computational overhead.

A least cost tree routing method might work in smaller networks where the cost of maintaining full network knowledge in each router is not too high. In larger networks or even on an Internet-wide scale this approach is unfeasible due to the communications and processing overhead involved in maintaining a full network graph and establishing or maintaining a Steiner tree for every (source, destination) area combination.

Another problem is the before-mentioned computational overhead of the Steiner tree. The Steiner tree problem is NP-complete \cite{KouMarkowskyBerman1981}, and the cost grows exponentially with the number of routers in a network.

The downsides of the Steiner tree make it unfeasible for larger networks. A shortest path tree (a tree consisting of all shortest paths from the source to all destinations) from the source to the destination area does not have these limitations. A single shortest path can be computed using a distance vector algorithm that does not require full network knowledge and has significantly less computational overhead compared to a Steiner tree.

For multicast it has been shown that shortest path trees can be preferable to Steiner trees in both fixed \cite{doar1993bad} and wireless ad-hoc \cite{NguyenXu2007} networks.
In our previous work, we have evaluated which type of routing tree would be most efficient, specifically for the geocast scenario \cite{MeijerinkBaratchiHeijenk2017}. We have shown that a shortest path tree has minimal additional cost in overall link usage compared to a perfect Steiner tree in a situation where destinations are geographically close.

\section{Algorithm Design}
\label{sec:algorithms} %TODO Dit aanpassen om de werkelijke structuur weer te geven
In this section, we will describe the process we followed to design our geographic routing algorithms. We will start by describing the path notation we will use in the rest of the paper. We will then discuss the simplest algorithm possible that will achieve our stated goal: flooding. In the subsections following that we will add conditions to build increasingly complex forwarding rules, resulting in our distance vector based algorithm. Following that, we will briefly analyze the performance of this algorithm. We continue by describing our path based forwarding algorithm, followed by short sections on possible link state approaches and hierarchical routing.

We define the primary goal for our geographic routing algorithm as follows: to deliver a message addressed to a certain area to all devices that cover that area with minimal cost. We will use the hop count (which for a tree we define as the total amount of transmission used per packet to reach all destinations) as our cost metric for simplicity, with a lower number of hops being better. 
To achieve our goals we choose to use a shortest path tree from the source to all routers that cover (advertise) the destination area. We also have the secondary design goals of limiting the processing overhead and using a system where no per destination signaling is needed.
Our algorithms are designed around the assumptions that all links in the network are symmetrical in both connectivity and cost.

\subsection{Path Notation}
We will use paths in the network to better explain and eventually build our routing protocol on. We define a shortest path $p_{n,m}$ through the network from a node $n$ to another node $m$.
\begin{gather*}
p_{n,m} = n \rightarrow \ldots \rightarrow m
\end{gather*}
We define the length of a path $l = |p_{n,m}|-1$ as the number of nodes it contains minus one. A path has a minimum length of 1 as $|p_{n,m}| \geq 2$  (assuming $n \neq m$), and can have an arbitrary number of nodes ($x_{a}, x_{b}, ...$ where $x_{a} \neq x_{b}$, $x\neq n$ and $x\neq m$) between $n$ and $m$. We define the k\textsuperscript{th} node in such a path as $p_{n,m}(k)$. For example, $p_{n,m}(1) = n$ in the path above. Note that we will use paths in the description of our distance vector approaches, even though this approach only uses cost information. The path information is always limited to the next and previous hop, information that would also be available to a purely distance vector based algorithm as it can be inferred from the cost information.

\subsection{Flooding} %TODO Deze misschienw eglaten, of alleen kort noemen in begin sectie....
\label{sec:algorithms:flooding}
The most straightforward approach to geocast is to simply flood the network with all traffic. This will lead to each packet traversing each link in the network at least once.

Flooding would guarantee that packets are delivered to all addressed destinations but there would be significant overhead, especially in larger networks with few router in the addressed destination. The total per packet transmission cost would be constant, equal to the number of links in the network assuming routers would ignore duplicate packets coming in on different links.

The transmission overhead is of course very large for such an approach, each link in the network would transmit the packet at least once. The processing overhead is limited to checking if a certain packet has already been processed before, or checking if an incoming packet has been received on the shortest path link to the source, provided that unicast routing info is present.

\subsection{Distance Vector}
\label{sec:algorithms:distance_vector}
Using every link in the network is not very efficient; we would like to construct a perfect shortest path tree through the network.
We can improve on the flooding protocol by introducing shortest path knowledge to the routers using a distance vector approach. A simple distance vector algorithm would give all routers knowledge of the shortest path next hop to all other routers. Coupled with geographic coverage information for these routers, this could enable geocasting in a network.

For the following algorithms we assume the routers have the following knowledge:
\begin{itemize}
	\item Coverage area for every router in the network.
	\item The cost and next hop for reaching every other router in the network.
\end{itemize}

Routers receive packets that have a geocast address in their destination field, as described in Section \ref{sec:prev_work:geo_addressing}. They check this address against the coverage area of the route advertisements they have received. Packets are forwarded to the routers that have overlapping coverage with the destination area.
We will now describe 4 distance vector algorithms in order of increasing complexity that use only the cost to other routers to forward packets to their destinations.

\begin{figure*}[t]%
	\centering
	\subfloat[Algorithm 1 \label{fig:algorithm_1} ]
	{{\includegraphics[width=0.25\linewidth, clip=true, trim=55mm 45mm 70mm 55mm]{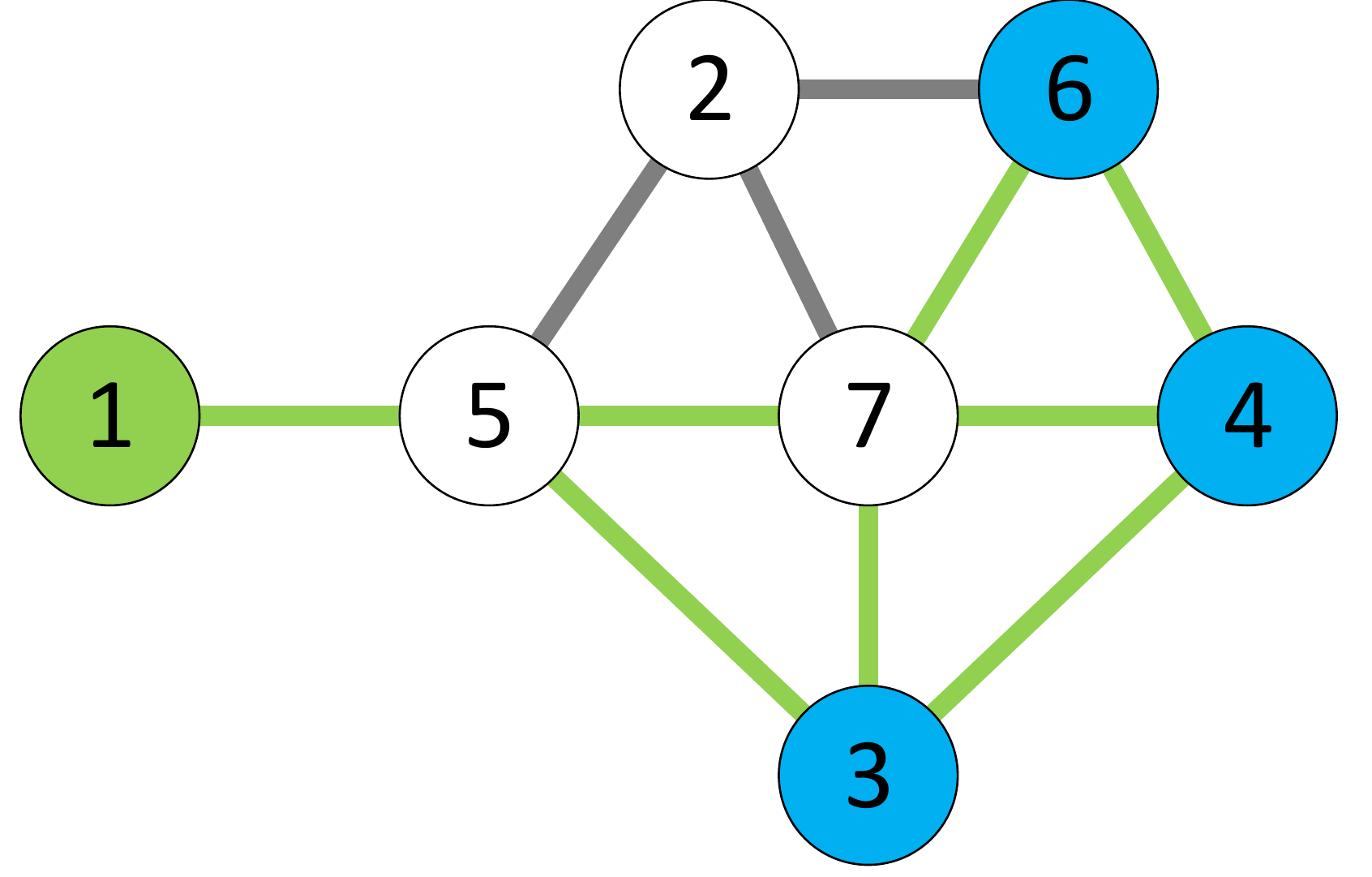} }}%
	%\qquad
	\subfloat[Algorithm 2 \label{fig:algorithm_2} ]
	{{\includegraphics[width=0.25\linewidth, clip=true, trim=55mm 45mm 70mm 55mm]{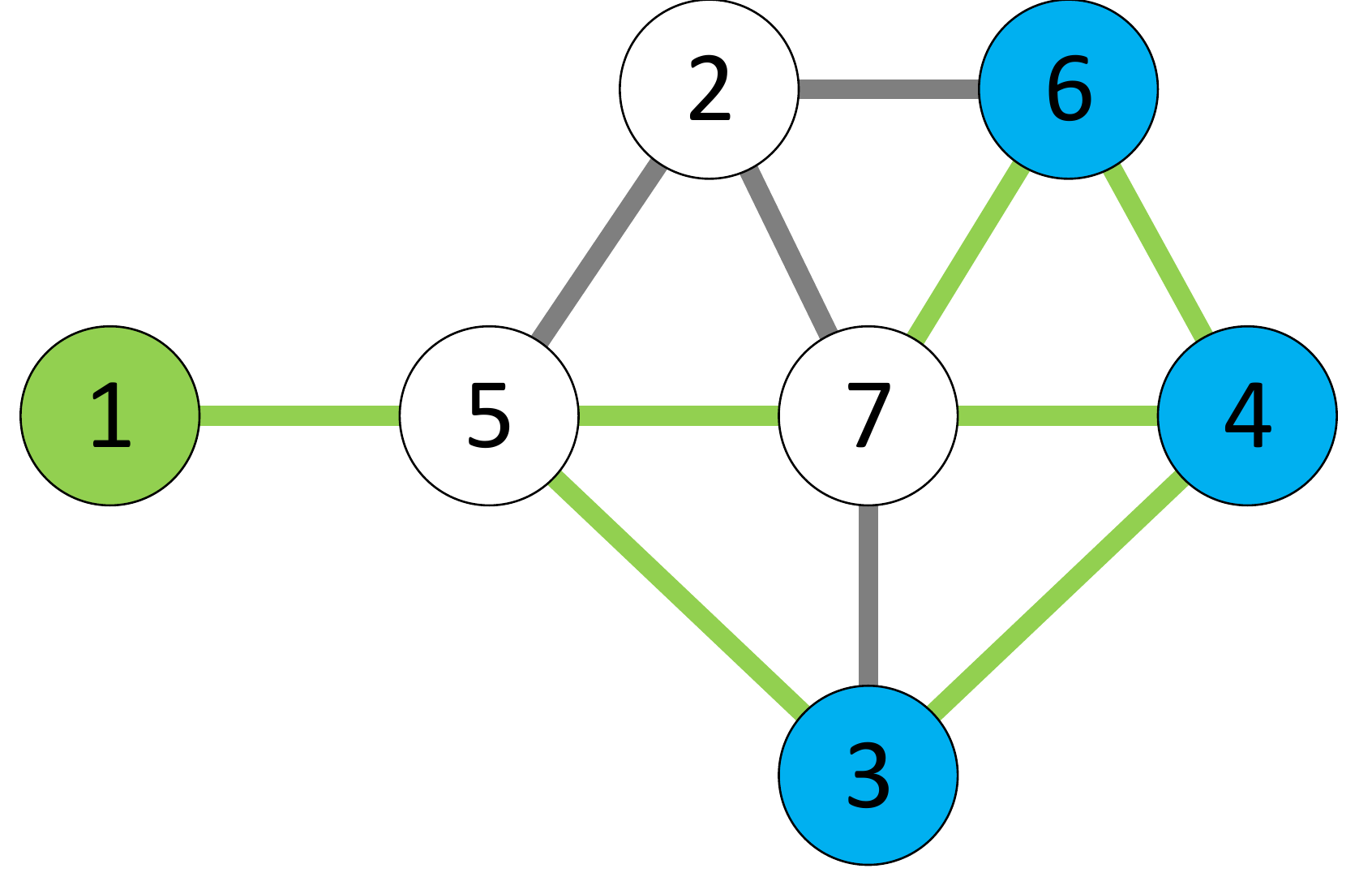} }}%
	%\qquad
	\subfloat[Algorithm 3 \label{fig:algorithm_3} ]
	{{\includegraphics[width=0.25\linewidth, clip=true, trim=55mm 45mm 70mm 55mm]{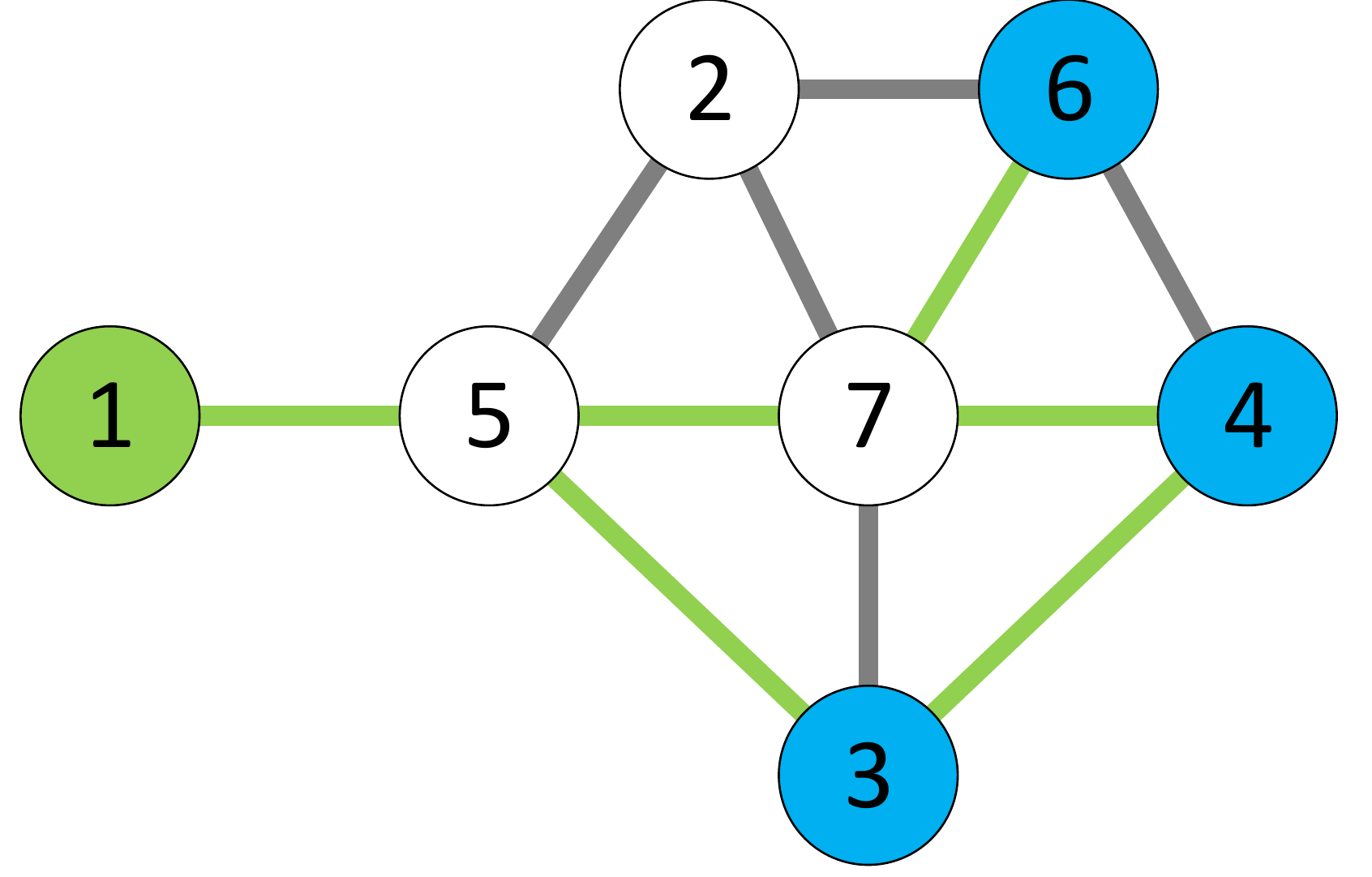} }}%
	%\qquad
	\subfloat[Algorithm 4 \label{fig:algorithm_4} ]
	{{\includegraphics[width=0.25\linewidth, clip=true, trim=55mm 45mm 70mm 55mm]{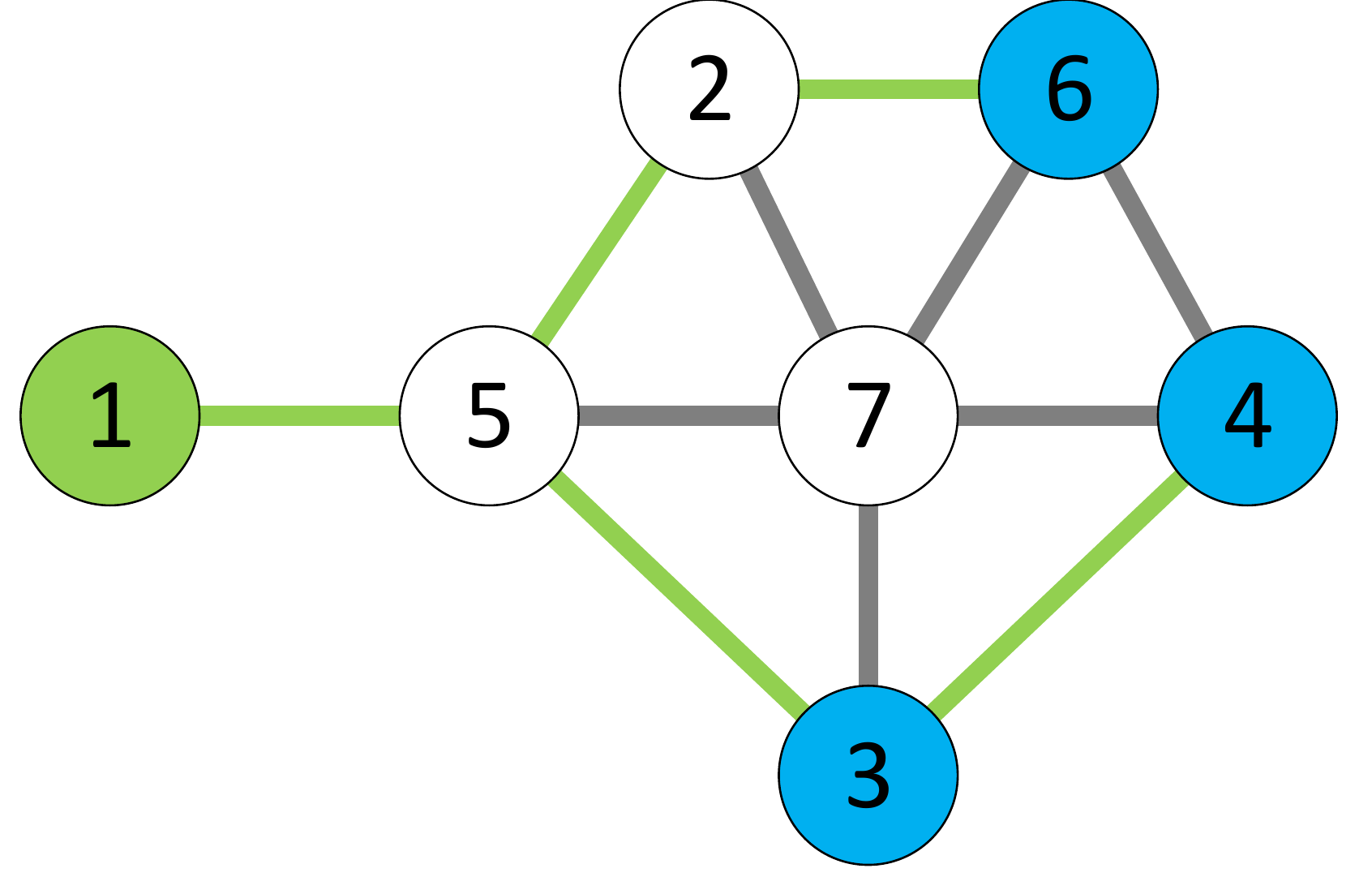} }}%
	\caption{Example routing tree of different distance vector algorithms}%
	\label{fig:algorithms}%
\end{figure*}

\subsubsection{DV Algorithm 1}
For our first attempt, we simply try to limit the flooding in the network to the `direction' of the destinations.
We use the term direction loosely here, as the actual geographical location of links and routers does not necessarily correspond to the area they cover.
In this simple approach, each router will forward packets it receives on its shortest path link to each of the destinations, except for the link the packet was received on. 

\begin{gather*}
fn(n,D)=\forall m \in neighbors(n) : \exists dst \in D : m = p_{n,dst}(2)
\end{gather*}	

The path $p_{n,dst}$ is the shortest path from the current router $n$ to a single destination $dst$. By definition this path passes through a next hop $m$ in the position $p_{n,dst}(2)$, that may be the same as the destination $dst$ (in which case the path would have a length of 1). With distance vector information each router is aware of at least two routers on such a path, itself, the destination and the next hop (which could be the same as the destination). The next hop router is simply the router that advertises the destination with the lowest cost (hops).

Each node on receiving a packet needs to evaluate if it forwards a packet based on a forwarding function $fn(n,D)$, with the destination set $D$ and the router itself $n$ as input that outputs the set of next hop nodes a packet with destination set $D$ should be forwarded to.
For each router $n$ that receives the packet we choose neighbors $m$ to forward to based on if they are the second entry on the known shortest path to a destination $dst$ in the destination set $D$ of the packet.

While this simple distance vector approach leads to a shortest path in the case of a single destination, with multiple destinations the performance is worse. As routers cannot know how they fit on a shortest path tree from the source to each destination, forwarding on the best next hop to all destinations would act like a form of limited flooding. This is caused by each router forwarding the packet on all its shortest path links to the destinations. While the algorithm floods the packet in the general direction of the destinations, there is still a large overhead compared to a shortest path tree.

In Figure \ref{fig:algorithm_1} we can see an example network consisting of 7 routers with a source router 1, and destination routers 3, 4 and 6. The links used by our simple algorithm are colored green. We can clearly see the `limited flooding' effect here, especially in router 7. This router is also forwarding to router 3 as it is the shortest path from the point of view of router 7, it is, however, not on the actual shortest path from the source to that destination and router 3 has already received the packet from another router.

\subsubsection{DV Algorithm 2}
It is obvious that the algorithm we described previously is not very efficient, as it uses more links than necessary to reach all destinations. We can improve the performance of the algorithm by ignoring packets that do not arrive on the reverse path interface to the source. Meaning that the shortest path $p_{n,src}$ from the current router $n$ to the $src$ has the previous hop, $ph$, as the second entry. This reverse path check already slightly reduces the average link usage due to not forwarding packets for which it is not on the reverse path but the `limited flooding' problem remains. Routers that are on the reverse path to the source and destination routers will still forward the packet to the other destinations in most cases.

To solve this forwarding problem we add a check for the cost towards the destination. We only forward if this cost reported by the previous hop is greater than the current routers cost. This checks that the current router $n$ is actually closer to the destination $dst$ than the previous router $ph$.

We extend our forwarding function $fn$ with these extra checks. This gives us the function $fn(n,D,src,ph)$, where we add the source $src$ and previous hop $ph$ of the packet as extra inputs. The output is the set of forwarding next hops as before.

\[
	fn(\bullet)=
	\begin{cases}
		\forall m \in neighbors(n) : \\
		\exists dst \in D : m = p_{n,dst}(2) \wedge & \text{if } ph = p_{n,src}(2) \\
		|p_{n,dst}| < |p_{ph,dst}| \\
		\\
		\emptyset & \text{if } ph \neq p_{n,src}(2)
	\end{cases}
\]

In Figure \ref{fig:algorithm_2} we can see the effect. Router 7 sees that the cost for the previous hop (router 5) to reach router 3 is 1. The cost for router 7 to reach router 3 is also 1, thus the packet is not forwarded on that link. Note that in this case we might actually prevent two transmission, as the packet might pass each other on that link as router 3 might send a packet for router 6 through 7. We still see router 4 sending a packet it receives from 3 to 6 as the cost 3 reports is 2 while router 4's own cost to 6 is 1.

\begin{figure*}[t]%
	\centering
	\subfloat[Binned performance \label{fig:dv_perf_lines} ]
	{{\includegraphics[width=0.46\linewidth, clip=true]{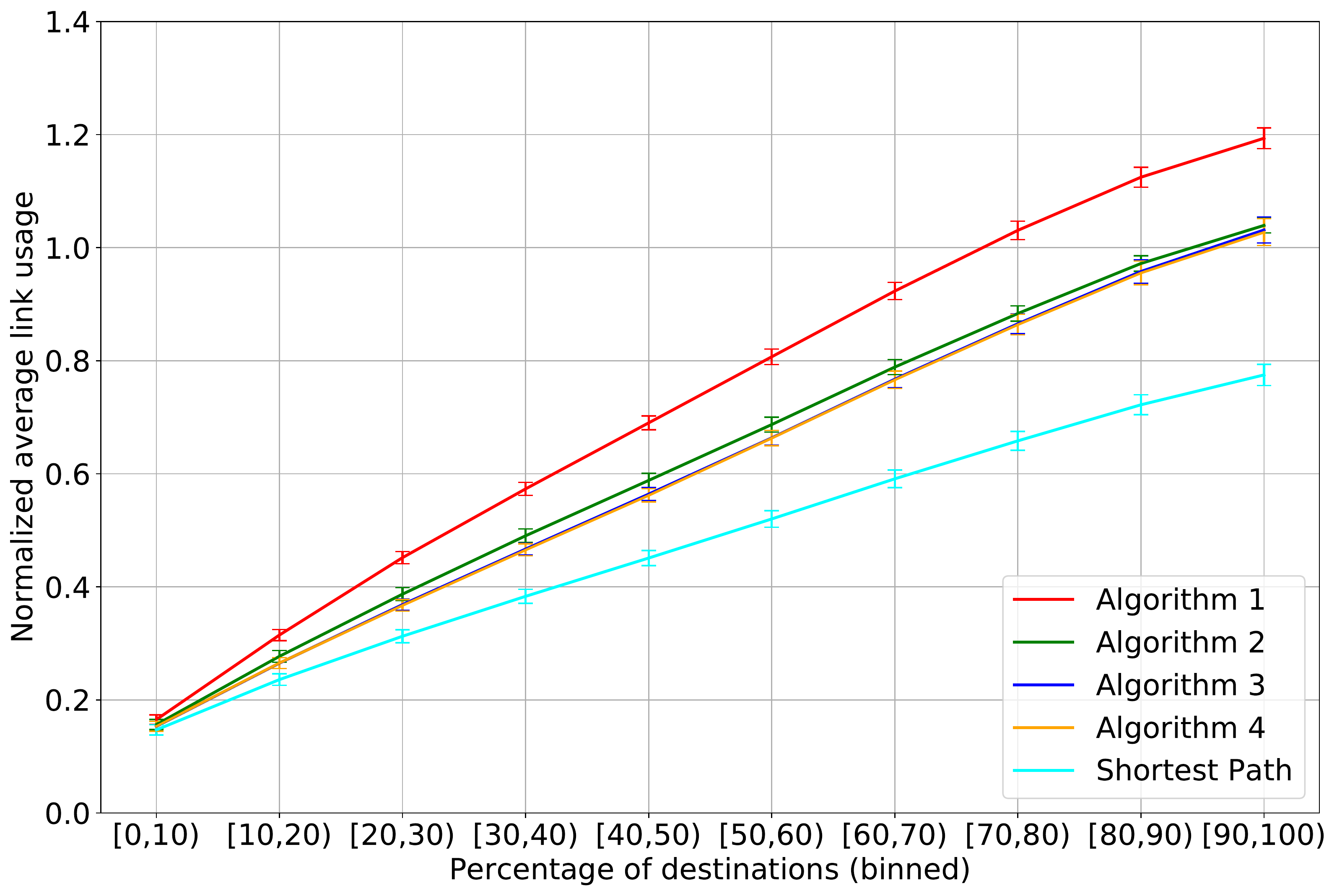} }}%
	\qquad
	\subfloat[Overall performance \label{fig:dv_perf_box} ]
	{\raisebox{0.038\height}{\includegraphics[width=0.46\linewidth, clip=true]{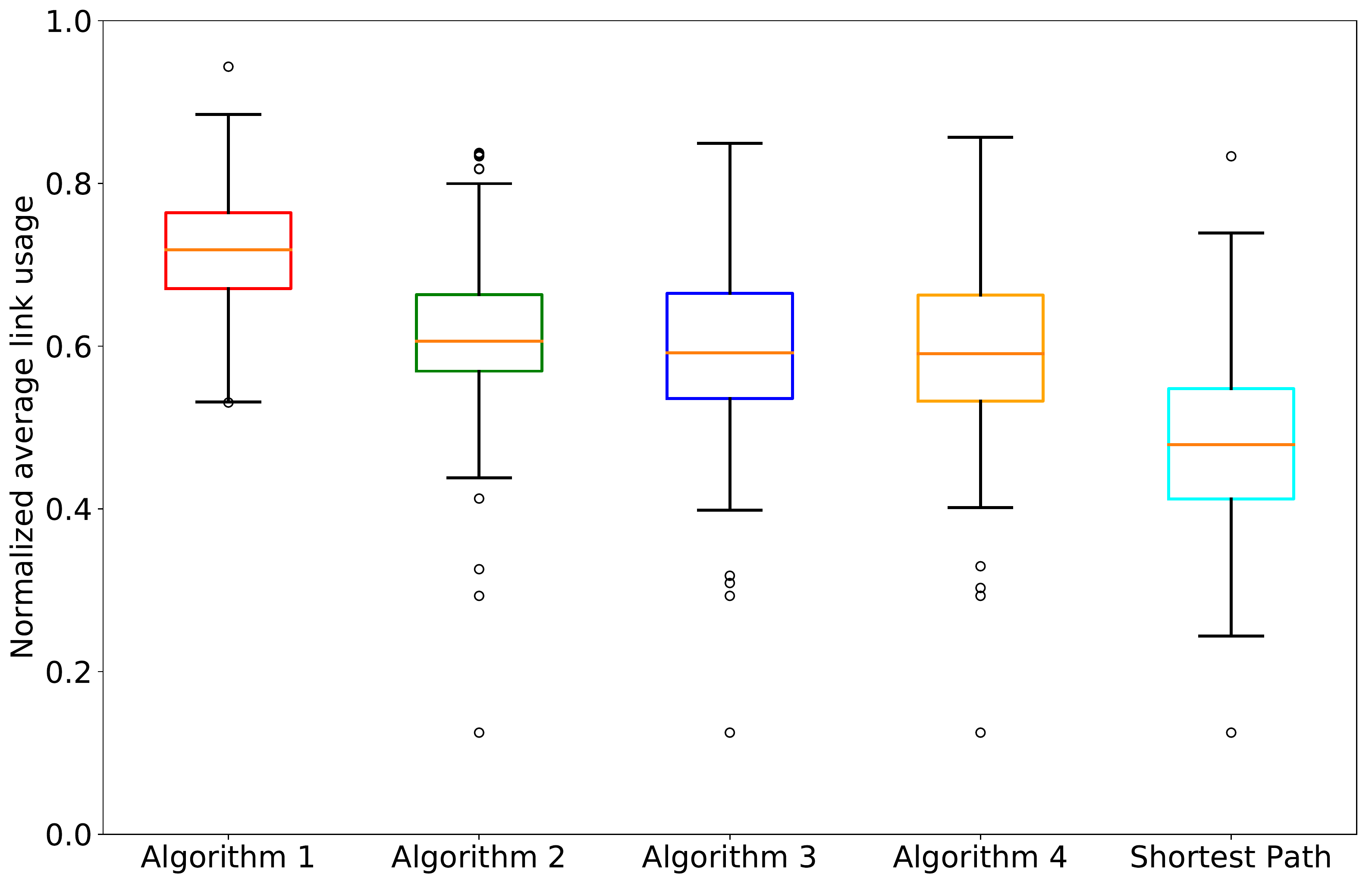} }}%
	\caption{DV algorithm performance}%
	\label{fig:dv_performance}%
\end{figure*}

\subsubsection{DV Algorithm 3}
Adding a check if the cost to reach the packet source through the current node is actually higher than the cost reported by the candidate next hop router for the source further improves the performance of the algorithm. Logically, if this was not the case, the candidate next hop should have already received this packet via another path. This prevents the packet from propagating `backwards' in certain situations. 
This check improves performance due to the fact that a router receives a packet because it is on the shortest path tree for at least one destination, but evaluates its forwarding for all destinations.
The `directed flooding' effect is reduced, but unneeded transmissions are not completely eliminated.

We once again extend out formula $fn$ to also take the next hop $nh$ as input leading us to the following function,

\[
	fn(\bullet)=
	\begin{cases}
		\forall m \in neighbors(n) : \\
		\exists dst \in D : m = p_{n,dst}(2) :\\
		|p_{n,dst}| < |p_{ph,dst}| \wedge & \text{if } ph = p_{n,src}(2) \\ 
		|p_{n,src}| > |p_{m,src}| \\
		\\
		\emptyset & \text{if } ph \neq p_{n,src}(2)
	\end{cases}
\]

with the check for the path length from candidate next hop $m$ to the source added. We can see the improvement of this addition in Figure \ref{fig:algorithm_3}, where the packets router 4 receives from router 3 and 7 are not forwarded to router 6 as the cost for router 4 to reach router 1 is identical to the cost of router 6 to reach router 1.

\subsubsection{DV Algorithm 4}
\label{sec:algorithm4}
Our final improvement to this algorithm is to prevent selecting the forwarding hop randomly when there are two equal length forwarding paths for a destination router. The path is now selected through a deterministic method. A router will select the next hop router based on its router ID. We chose (arbitrarily) to use the lowest next hop ID for this. 
This choice forced packets down the same links when there is a choice. This change will force packets with similar destinations over the same link, leading to a more optimal tree due to less overall link usage. Our path based algorithm will exploit this deterministic behavior for its forwarding as we will explain in Section \ref{sec:algorithms:path_distance_vector}.

%TODO dit uitbreiden met meer uitleg... maar hoe?
The results of this modification are visible in Figure \ref{fig:algorithm_4}. Note that had we chosen to use the highest next hop id, the forwarding tree would be equal to that shown in Figure \ref{fig:algorithm_3}, so this modification does not always make the algorithm more optimal. It does however force packets over somewhat similar paths in cases where shortest paths to multiple destinations share a similar low next hop id.

\subsection{Intermezzo: DV Performance}
After evaluating the performance of the different distance vector algorithms we see that none come close to the performance of the shortest path tree. In some cases there are even two identical packets traversing the same link when routers forward at the same time, resulting in packets `crossing' each other on the link. However, these distance vector approaches do have the benefit of having very low computational overhead. They rely on the transmission of coverage information per router in addiction to the `normal' distance vector information associated with such a protocol.

In Figures \ref{fig:dv_perf_lines} and \ref{fig:dv_perf_box} we compare the performance of the algorithms described above and the performance a shortest path tree (cyan line/box) would have. As our main metric we use the average link cost of the algorithm; we define this as the average number of transmission used to reach a certain number of destinations. The error bars in Figure \ref{fig:dv_perf_lines} show the 95\% confidence interval for that bin.

We run our evaluation over a large set of real world network graphs taken from the Topology Zoo \cite{topzoo}. Our exact evaluation method will be explained in Section \ref{sec:evaluation}, for now we will present the performance of an algorithm as a normalized link usage metric. This is calculated by taking the average link usage for a certain number of destinations in a network and dividing it by the number of links in the network. A value of 1 represents all links being used, a value above 1 shows that one or more links are used multiple times. We then bin the number of destinations per 10\% and average these numbers over all networks with more than 10 nodes. The results are shown in Figure \ref{fig:dv_perf_lines}. Figure \ref{fig:dv_perf_box} shows box-plots for all algorithms averaged over all numbers of destinations. Using this plot we can compare the overall performance of the different algorithms. We can mainly see that the largest improvement was made with relatively simple additions to our forwarding rules, and that later additions only marginally improve the link cost of the forwarding tree.

We can see that the improvement between our first and second distance vector algorithms is relatively large, while further improvements provide only minor benefits. Overall, the link usage of our 4th distance vector algorithm is around $12\%$ larger than the optimal shortest path tree link usage for a small number of destination to around $32\%$ when (almost) all routers are addressed. On average the link usage is $24\%$ larger than our shortest path tree target.
This result implies that for situations where only a small amount of routers in the network would be addressed, the simple solution might be viable, but for large destination sets the overhead is large.

\subsection{Path Based Distance Vector}
\label{sec:algorithms:path_distance_vector}
While the performance of the distance vector algorithms presented thus far is optimal in the case of the network presented in Figure \ref{fig:algorithms}, this does not hold for larger networks as we can see by looking at the link usage in Figures \ref{fig:dv_perf_lines} and \ref{fig:dv_perf_box}. The main problem with the distance vector approach is that routers have no information of how they fit in the complete forwarding tree in the network. Considering the limited knowledge that is used to calculate forwarding decisions in the DV algorithms we can certainly do better with more information about other paths. As stated before, our aim is to establish a forwarding tree which is as close as possible to the shortest path tree.
To prevent the limited flooding effect and also keep the amount of information that needs to be distributed in the network low, we have investigated an approach where routers not only know the next hop to each destination, but also know the complete path to other routers, somewhat like the Border Gateway Protocol (BGP) \cite{RekhterLiHares2005}.
This information would allow routers to make decisions that can lead to a close to optimal shortest path tree, at the cost of computational overhead and larger route advertisements.

Our proposed path based algorithm evaluates forwarding decisions on a per destination router basis. The destination address of a packet arriving at a router is mapped to a set of routers advertising (partial) coverage of this destination. A router can now use its shortest path information for each of these destinations to evaluate its forwarding options, similar to the purely cost based algorithms presented earlier.

We develop our path based algorithm on the basis of the 4th distance vector algorithm we presented earlier. The forwarding rules from this previous algorithm are extended to no longer use only the number of hops but the entire path  to base the evaluation on.

The main problem we try to fix with this path-based approach is that routers have no knowledge of alternate paths through the network while making their forwarding decisions. This can lead to extra transmissions in some cases where for example destination routers think they need to forward the packet to other destinations, while in reality these have already been reached. We attempt to solve this problem by keeping track of two distinct paths towards each other node in the graph when possible. We will describe our route distribution method later in the paper. For now we will assume each router knows the one or two (when such an alternate path exists) shortest paths towards all other routers.

We will start with explaining our path distribution method. The algorithm calculating the next hop(s) will be described following this, followed by an explanation of the lowest next hop ID rule. We conclude this subsection by describing situations in which the algorithm fails to deliver an optimal shortest path tree.

\subsubsection{Route distribution}
\label{sec:algorithms:path_distance_vector:route_distribution}
Each router or node in the network advertises its coverage area(s) on each of its links. This advertisement contains the path, initially only the advertising router. A router must append its own id to the path it is propagating.

An advertisement packet contains the $Area$, $Cost$ and $Path$ to reach it. The $Path$ is a set of router IDs: $Path=ID_{0}, ID_{1}, ... , ID_{n}$, where $ID_0$ is the advertising router and $ID_n$ the previous hop as seen from the receiving router.

The advertising router id combined with the coverage area should be unique in the network. Using this method, different routers with identical or overlapping coverage areas can be uniquely identified. This allows a router to know which of its links leads to routers that cover the geographic area in the destination field.

A router will transmit the best paths it knows to all neighboring routers known to it on each of its links except if the router on the other end is contained in a path. In that case the router will transmit an alternative, possibly longer, path that does not contain the other router. If an alternate route is not available, for example because the router has no other links, the path containing the neighbor is returned. A neighbor can detect such a loop due to the path information in the advertisement. This system with two distinct paths ensures that routers have knowledge about the existence or nonexistence of alternate routes.

\begin{figure}[t]
	\centering
	\includegraphics[width=1.0\linewidth, clip=true, trim=17mm 20mm 7mm 42mm]{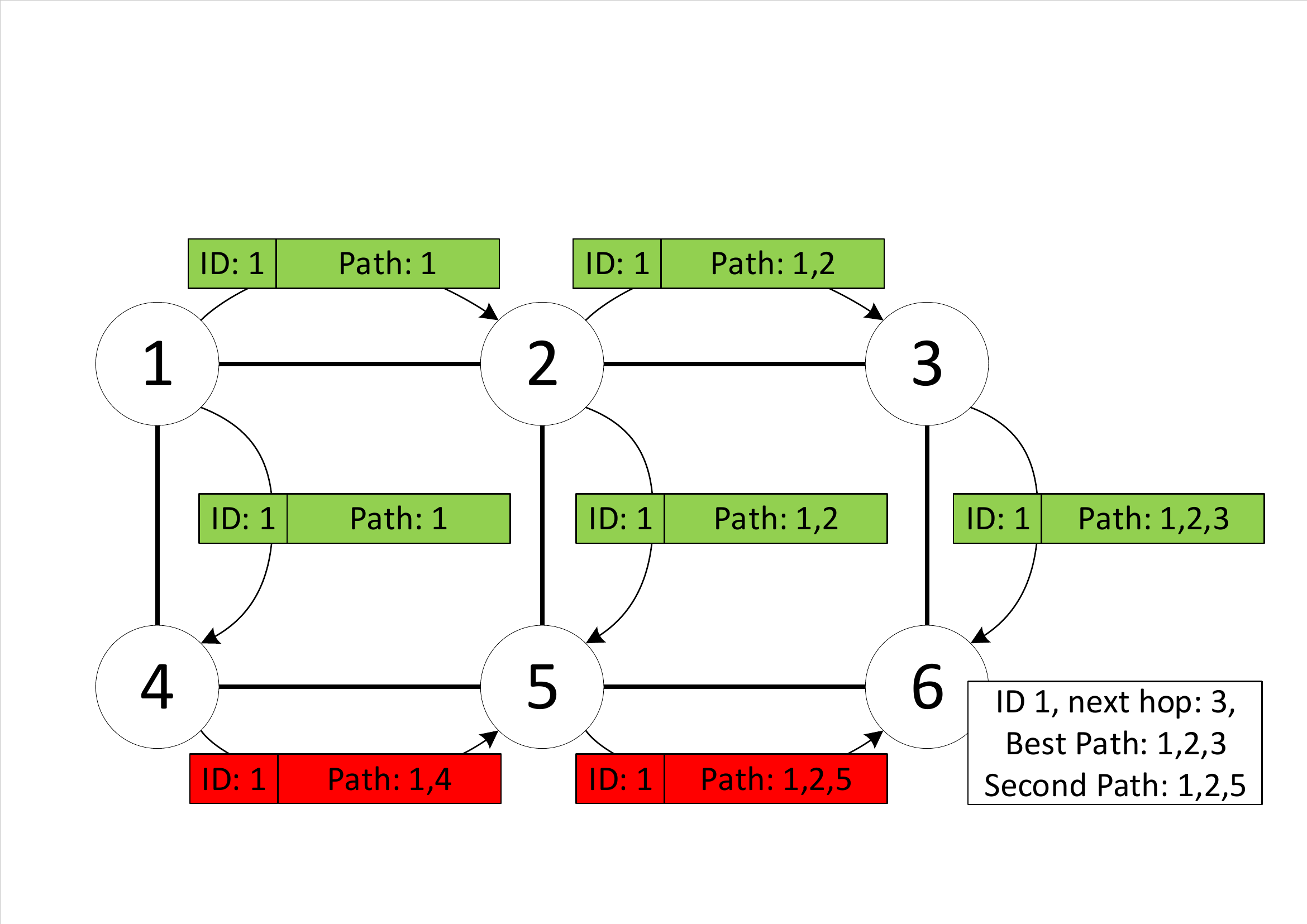}
	\caption{Example of route distribution}
	\label{fig:route_distribution}
\end{figure}

Figure \ref{fig:route_distribution} shows an example of the path for the router with id $1$ being distributed through a network with 6 nodes. The paths marked in green are chosen by the receiving routers due to the lowest next hop id rule. The paths marked in red are kept as second best path by the receiving routers.

A router will need to keep track of the advertisements it receives on all its links. Assuming each router has one area it will cover (this could also be zero or multiple areas), there will be an entry per destination per link resulting in $D \times deg(n)$ entries for a router where $D$ is the number of destinations and $deg(n)$ the node degree of the router itself (the number of links this router has). The entire network will have $\sum_{n \in N} deg(n) \times D^{2}$ entries. 

The resulting routing table has a number of entries that is at least equal to the number of distinct (router, coverage area) pairs in the network. The worst-case scenario is that the number of entries is twice the number of pairs due to the existence of alternate routes. An example would be router 3 in Figure \ref{fig:route_distribution}, which will receive an alternate route from router 6 with the path $1 \rightarrow 2 \rightarrow 5 \rightarrow 6$, as the best route known to router 6 passes through router 3 itself.

In the event a router detects a link as no longer available, by no longer receiving advertisements on that link, it will stop advertising paths that contain this link to its neighbors. As routers do not propagate paths that contain themselves in the path, eventually all nodes will have updated path information.

\subsubsection{Choosing forwarding next hops}
\label{sec:algorithms:path_distance_vector:forwarding}
Once a router receives a packet it will evaluate the packet's source and destination. The destination geocast address is translated to all known routers in the network that (partially) cover that area using the method described in our previous paper \cite{meijerink2016efficient}. After the router generates this list of the covering router ids it can move on to the forwarding step.

A router needs to evaluate each (source, destination) combination using a simplified $fn$ function which takes the current router $n$, destination set $D$, source $src$ and previous hop $ph$ as input values.
\begin{align*}
	fn(n,D,src,ph) = &\exists m \in neighbors(n) : \\
	&\forall dst \in D : dst \neq n \wedge \\
	&m = p_{n,dst}(2) \wedge m \neq ph
\end{align*}
A given destination $dst$ from the set of destinations $D$ and the next hop $m$ to that destination are evaluated for forwarding if the current router $n$ is not the destination, the next hop for the destination is not the previous hop $ph$ and we do not already have the next hop in the forwarding hop list because of another destination in the set. This initial simple step can be seen in Algorithm \ref{alg:lookup}. We now compare three possible paths to each other. We check the shortest path from source to destination through the current router, the shortest path as seen from the previous hop and the shortest path as seen from the candidate next hop for each candidate next hop that passes the initial checks.

\begin{algorithm}[b]
	\capstart
	\LinesNumbered 
	\setstretch{0.7}
	\SetAlgoLined
	\SetKwInOut{Input}{Input}
	\SetKwInOut{Output}{Output}
	\Input{Destination routers list \textit{D}\\Previous hop \textit{prev\_hop}\\Source router \textit{src} }
	\Output{List $result$ with next hops for the packet}
	List $result$\tcp*[r]{Initialize list}
	\tcp{Check if we are entry router}
	\If{prev\_hop == -1}{ 
		\ForEach{dst $\in$ D}{
			\If{self.nextHop(dst) $\notin$ result}{result.add(self.nextHop(dst))}
		}
	}\Else{
		\ForEach{dst $\in$ D}{
			$nextHop \leftarrow self.nextHop(dst)$\;
			\If{$dst != self$ and $nextHop != prev\_hop$ and $nextHop \notin result$ and $find\_dif(dst, src, prev\_hop, nextHop)$}{
				result.add(nextHop)
			}
		}
	}	
	\caption{Next hop lookup algorithm for a router \label{alg:lookup}}
\end{algorithm}

\begin{algorithm}[b!]
	\capstart	
	\LinesNumbered 
	\setstretch{0.7}
	\SetAlgoLined
	\SetKwInOut{Input}{Input}
	\SetKwInOut{Output}{Output}
	\Input{ Candidate next hop \textit{next\_hop}\\The previous hop \textit{prev\_hop}\\ Destination \textit{dst}\\ Source \textit{scr} }
	\Output{Boolean $result$}
	
	$nh\_src$ = self.pathTo($next\_hop, src$)\;
	$nh\_dst$ = self.pathTo($next\_hop, dst$)\;
	$ph\_src$ = self.pathTo($prev\_hop, src$)\;
	$ph\_dst$ = self.pathTo($prev\_hop, dst$)\;
	
	$ph\_dst\_removed$ = $ph\_dst$\;
	\ForEach{$n \in ph\_src$}{
		\If{$n \in ph\_dst\_removed$}{$ph\_dst\_removed$.remove($n$)}
	}
	
	$ph\_src\_removed$ = $ph\_src$\tcp*[r]{Copy list}
	\ForEach{$n \in ph\_dst$}{
		\If{$n \in ph\_src\_removed$}{$ph\_src\_removed$.remove($n$)}
	}
	$ph\_src\_dst$ = $ph\_src\_removed$\tcp*[r]{Copy list}
	$ph\_src\_dst$ += $ph\_dst\_removed$ \tcp*[r]{Add lists}
	
	$nh\_src\_removed$ = $nh\_src$\tcp*[r]{Copy listp}
	\ForEach{$n \in nh\_dst$}{
		\If{$n \in nh\_src\_removed$}{$nh\_src\_removed$.remove($n$)}
	}
	$nh\_dst\_removed$ = $nh\_dst$\tcp*[r]{Copy list}
	\ForEach{$n \in nh\_src$}{
		\If{$n \in nh\_dst\_removed$}{$nh\_dst\_removed$.remove($n$)}
	}
	$nh\_src\_dst$ = $nh\_src\_removed$\tcp*[r]{Copy list}
	$nh\_src\_dst$ += $nh\_dst\_removed$ \tcp*[r]{Add lists}
	
	$p\_src\_dst$ = $ph_src$\;
	$p\_src\_dst$ += $nh\_dst$ \tcp*[r]{Path through this router}
	
	$result$ = false\;
	\If{length($ph\_src\_dst$ $\geq$ length($p\_src\_dst$))}{
		\If{self $\in nh\_src$}{
			$result$ = true\tcp*[r]{on nh to src}
		}\ElseIf{$length(nh\_src\_dst) > length(p\_src\_dst)$}{
			$result$ = true \tcp*[r]{Other path is worse}
		}\ElseIf{$length(nh\_src\_dst) == length(p\_src\_dst)$}{
			$temp$ = [self] + $ph\_src$\;
			\For{$i=1; i < length(nh\_src)+1); i++$}{
				\If{$nh\_src[length(nh\_src)-i] != temp[length(temp)-i]$ }{
					\If{$nh\_src[length(nh\_src)-i] > temp[length(temp)-1]$}{
						$result$ = true\;
					}					
					break\;
				}
			}
		}
	}
	return $result$\;
	
	\caption{Find path difference \label{alg:find_dif}}
\end{algorithm}

\begin{figure}[t]
	\centering
	\includegraphics[width=0.48\linewidth, clip=true, trim=54mm 23mm 87mm 37mm]{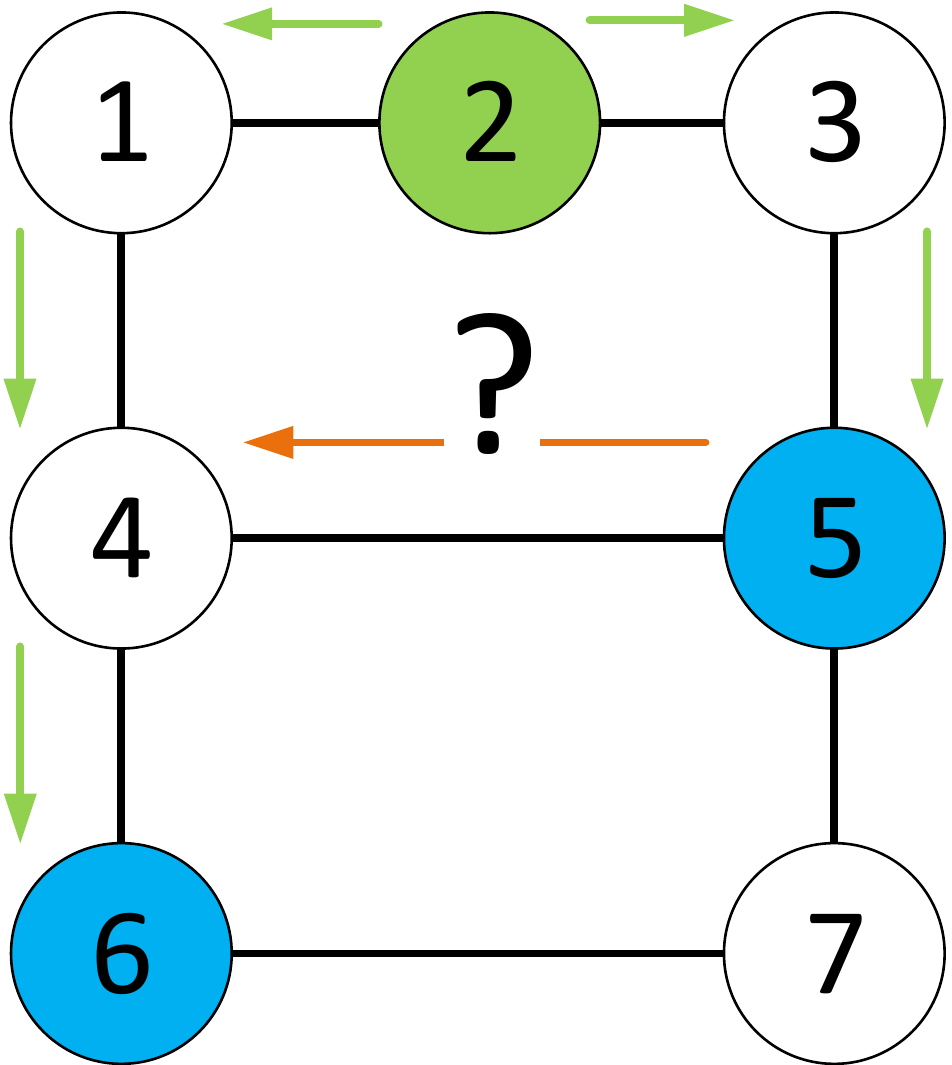}
	\caption{Forward lookup from node 5 to 4}
	\label{fig:diff_example}
\end{figure}

We will use Figure \ref{fig:diff_example} to illustrate our rules. In this figure, router 2 is the source for a packet that needs to be delivered to routers 5 and 6. We will focus on router 5 that needs to decide if it will forward the packet it received in the direction of router 6. There are two shortest paths from 5 to 6, namely $5 \rightarrow 4 \rightarrow 6$ and $5 \rightarrow 7 \rightarrow 6)$. Both candidate next hops pass the initial check described in Algorithm \ref{alg:lookup}. We will evaluate the path through router 4 for this example as it has a lower id, the exact reasoning behind this choice will be explained in Section \ref{sec:algorithms:path_distance_vector:nhid}.

We start by constructing the path from the source to the destination as seen from the next hop router $p^{nh}_{src,dst}$, that is the shortest from the source to the destination that the next hop can know of given the path information it has.
\begin{gather*}
	p^{nh}_{src,dst} = p_{src,nh} \triangle p_{nh,dst}
\end{gather*}
Here $p_{src,nh}$ is the shortest path from the \textit{source} to the \textit{next hop} node and $p_{nh,dst}$ the shortest path from the \textit{next hop} to the \textit{destination}. We define the $\triangle$ operation as the concatenation of the two paths excluding duplicate routers except the one that connects the paths. This is done in Algorithm \ref{alg:find_dif} on lines 15 to 24.

In our example $p_{src,nh} = 2 \rightarrow 1 \rightarrow 4$ and $p_{nh,dst} = 4 \rightarrow 6$ which gives us $p^{nh}_{src,dst} = 2 \rightarrow 1 \rightarrow 4 \rightarrow 6$. In some situations both paths could share multiple routers at the start, leading to the exclusion of all but the last of these shared routers in the constructed path. In our example there is only one shared router so it stays on the path as it is also the last.

We now construct a similar path as seen from the previous hop router $p^{ph}_{src,dst}$.
\begin{gather*}
	p^{ph}_{src,dst} = p_{src,ph} \triangle p_{ph,dst}
\end{gather*}
Where the path is constructed from the shortest path from the \textit{previous hop} to the \textit{source} $p_{src,ph}$ and the shortest path from the \textit{previous hop} to the \textit{destination} $p_{ph,dst}$. This is done in Algorithm \ref{alg:find_dif} on lines 5 to 14.

In our example $p_{src,ph} = 2 \rightarrow 3$ and $p_{ph,dst} = 3 \rightarrow 2 \rightarrow 1 \rightarrow 4 \rightarrow 6$ which gives us $p^{ph}_{src,dst} = 2 \rightarrow 1 \rightarrow 4 \rightarrow 6$. Note that the previous node reports a longer path towards node 6 to node 4 as the shorter path passes through node 4 itself.

Finally we construct the path that the packet will flow if we forward it $p^{n}_{src,dst}$, which is the path as seen from the current router $n$.
\begin{gather*}
	p^{n}_{src,dst} = p_{src,ph} \triangle n \triangle p_{nh,dst}
\end{gather*}
The path is constructed from the path from the \textit{previous hop} to the \textit{source}, the \textit{next hop} to the \textit{destination} and the router \textit{n} itself.

Using the example in Figure \ref{fig:diff_example}, this results in $p_{ph,src} = 3 \rightarrow 2$ and $p_{nh,dst} = 4 \rightarrow 6$ which gives us $p^{n}_{src,dst} = 2 \rightarrow 3 \rightarrow 5 \rightarrow 4 \rightarrow 6$. Note that we include the current node in this path!
Now that we have all relevant paths we can compare their lengths to each other.

\begin{align*}
pd(n,src,D,ph) = &\forall m \in fn(n,src,D,ph) :
\exists dst \in D :\\ &|p^{n}_{src,dst}| \leq |p^{ph}_{src,dst}| \wedge \\
&|p^{n}_{src,dst}| \leq |p^{m}_{src,dst}|
\end{align*}

\begin{figure}[t]
	\centering
	\includegraphics[width=0.8\linewidth, clip=true, trim=17mm 31mm 57mm 55mm]{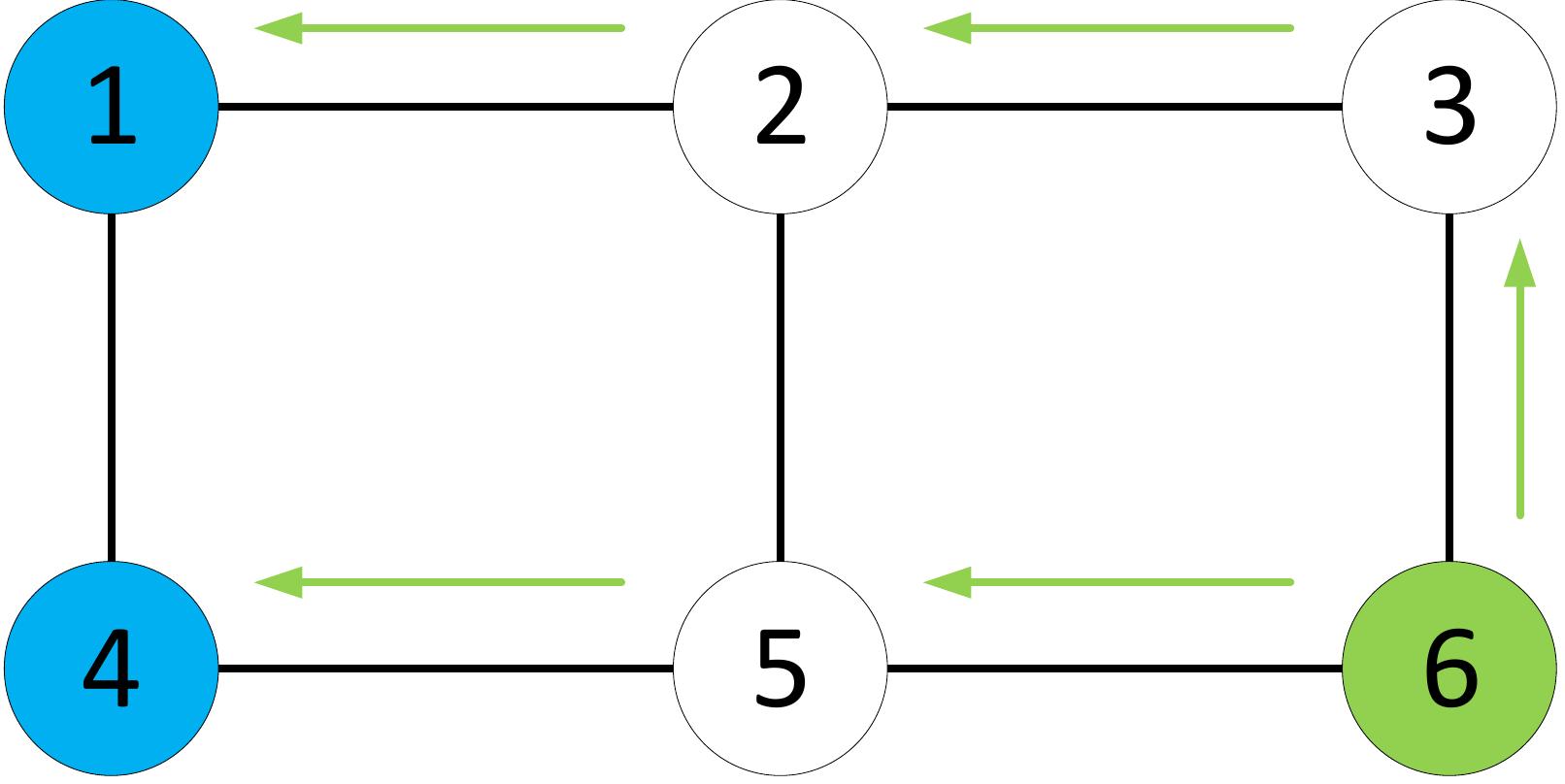}
	\caption{Forwarding from node 6 to nodes 1 and 4}
	\label{fig:forwarding_example}
\end{figure}

Where $pd$ takes the current router $n$, source $src$, destination set $D$ and the previous hop $ph$ and returns a set of forwarding next hops from the set of neighbors of $n$. The forwarding candidates are taken from the previous function $fn$, that already reduced the set of possible forwarding next hops.

If $n \in p^{nh}_{src,dst}$ we can skip these checks and always forward, as there is no known shorter path visible to the next hop router. A better path cannot exist in this case as the next hop router would report a path that does not contain us that is shorter.

In our example $|2 \rightarrow 3 \rightarrow 5 \rightarrow 4 \rightarrow 6| \leq |2\rightarrow 1 \rightarrow 4\rightarrow 6|$ which gives us ($5 \leq 4$), which is false. The result is that router 5 will not forward the packet to router 4.
If the statement would have been true the second evaluation would have also been False (in this case it is even the same path).

\begin{figure}[t]
	\centering
	\includegraphics[width=0.48\linewidth, clip=true, trim=54mm 20mm 87mm 41mm]{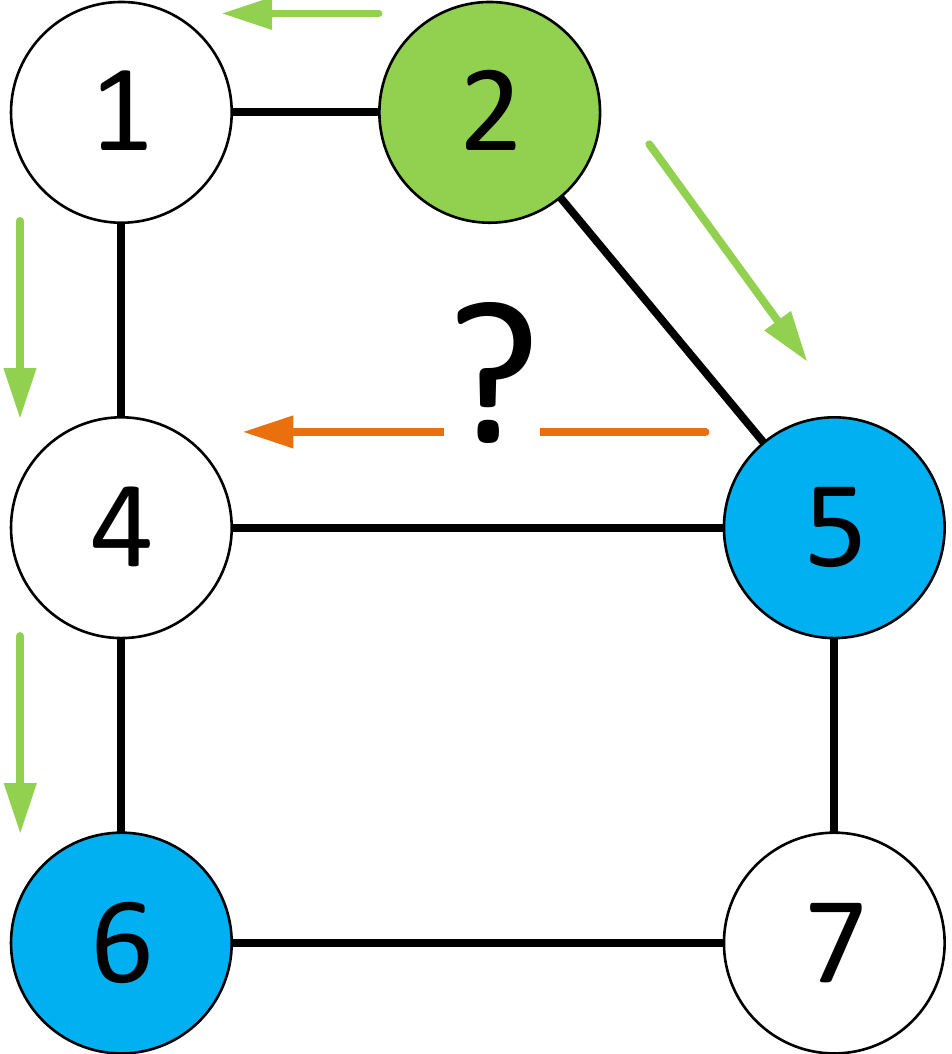}
	\caption{Forwarding lookup from node 5 to 4}
	\label{fig:diff_example2}
\end{figure}

Pseudo code for the entire forwarding operation are given in Algorithms \ref{alg:lookup} and \ref{alg:find_dif}. In Algorithm \ref{alg:lookup} we show the initial forwarding step. The router will always forward on the shortest path to the destinations if it is the entry router for the packet in the network (check against by $id == -1$). If the packet is received from another router in the same network the router finds a candidate next hop ($nextHop$), checks if it is not the destination, the packet is not returned on the previous hop and that the next hop is not already included in the forwarding next hops list. If all these checks pass the router will perform the forwarding check described above. The pseudo code for these steps is given in Algorithm \ref{alg:find_dif}. Using this approach we can achieve close to optimal shortest path trees in most networks, as we will show in Section \ref{sec:evaluation}.

\subsubsection{Lowest next hop ID}
\label{sec:algorithms:path_distance_vector:nhid}
When a router has the option of two or more paths of identical length to forward on for a certain destination it has a choice. We let routers base this choice on the lowest router id of the next hop. This makes the choice between paths deterministic and by extension allows routers further in the forwarding tree to know for which destination they are part of the forwarding tree.

Consider the network from Figure \ref{fig:forwarding_example}, here a tree is constructed from node 6 to nodes 1 and 4 based on the rules described previously. We choose router 3 as the forwarding hop to router 1 over router 5 because it has a lower id.
In Algorithm \ref{alg:find_dif} (which will be explained fully later on) we can see the lowest id rule implemented in lines 43 to 52.

A similar choice also needs to be made if a router needs to choose between forwarding or not based on path knowledge of its candidate next hop. The router can compare the two paths and check which of the paths has the lowest ID next hop from where they diverge from each other. Using this method a router can determine where it sits in the forwarding tree and for which destinations it should forward.

To illustrate this, we will use a modified version of the network shown in Figure \ref{fig:diff_example} with node 2 removed, shown in Figure \ref{fig:diff_example2}. The source is router 2 and the destination are routers 5 and 6 as before. Router 5 has to decide if it forwards the packet is has received to router 6. There are two possible paths available to the router $5 \rightarrow 4 \rightarrow 6$ and $5 \rightarrow 7 \rightarrow 6$, where the first path has the lower next hop ID. This path through router 5, $2 \rightarrow 5\rightarrow 4\rightarrow 6$, is now compared to the path as seen from the candidate next hop router 4: $2 \rightarrow 1 \rightarrow 4 \rightarrow 6$. These paths diverge from each other after router 2, where the best path as seen from router 4 has the lowest next hop ID, so router 5 will choose not to forward.

This method of choosing one forwarding path over the other allows our system to only use one path to each destination. In some cases such as the one shown in Figure \ref{fig:diff_example2} this leads to a forwarding tree that has a slightly higher cost than a Steiner tree, but never multiple paths are used to reach the same destinations.

\subsubsection{Close to optimal shortest path trees}
\label{sec:algorithms:path_distance_vector:sp_trees}
The algorithm using limited path knowledge described above constructs a close to optimal shortest path tree between a given source and set of destinations.

\begin{figure}[t]
	\centering
	\includegraphics[width=0.6\linewidth, clip=true, trim=70mm 30mm 60mm 30mm]{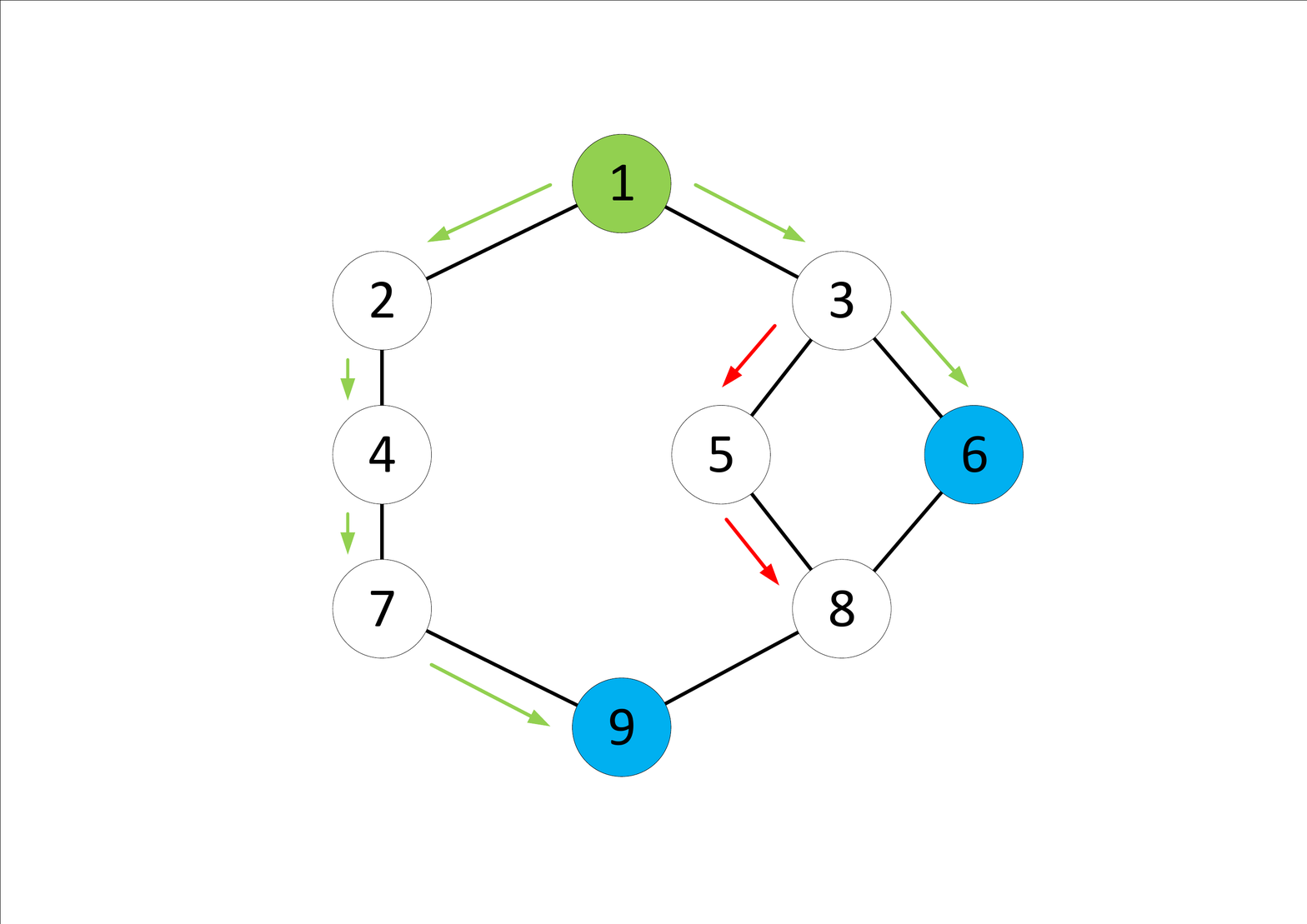}
	\caption{Shortest example of forwarding error loop}
	\label{fig:loop_example}
\end{figure}

However, the algorithm fails to construct an optimal shortest path tree in some specific situations where the network contains loops within loops. A minimal example of one such network can be seen in Figure \ref{fig:loop_example}. In this figure node 1 represents the source, nodes 6 and 9 are the destinations.

Routers in the small loop will receive advertisements for the source and destination from both their neighbors as they keep track of the two best paths (when multiple paths exists) to the source. Both paths will use the small loop to reach those destinations as the distance is shorter compared to the large loop. The result is that the routers inside the small loop have no knowledge of the alternate path through the larger loop and mistakingly believe they should forward the message for node 9. The router that connects the small loop to the larger loop (router 8 in Figure \ref{fig:loop_example}) does have this knowledge and will correctly not forward the packet.

The maximum cost of extra link usage is half the number of links on the small loop. This occurs in the case where the destination in the small loop is on the side of the higher next hop id from the source (like node 6 in Figure \ref{fig:loop_example}).

In practice, as our evaluations will show in Section \ref{sec:evaluation}, this problem almost never occurs and when it does the extra overhead caused is minimal.

\subsection{Link State}
\label{sec:algorithms:link_state}
For completeness we have to mention the option of using a link state algorithm for geographic routing.
Using a link state algorithm that can provide full network knowledge, each node can determine if it is on the shortest path between a given source and destination and base its forwarding decisions on that.

The benefit of this approach would be the perfect shortest path trees for any given (source, destinations) combination. The main downside would be computational, storage and communication overhead of the full network graph as compared to the other alternatives presented.

Link state algorithms would also allow each router to compute a Steiner tree for a given source and destination set, allowing the network in theory to forward using the most optimal tree possible. 

The amount of data that would need to be transmitted and the computational overhead for such an approach would be large and scale with the network. The main problem computation wise would be that the router has to calculate if it is on the shortest path from source to destination for every unique source destination combination.
Considering these drawbacks we think a link state approach is not feasible and will not consider it further in the remainder of this paper.
We will also show in Section \ref{sec:evaluation} that the path based distance vector approach already comes close to a shortest path tree.

\subsection{Hierarchical Routing}
Due to the ability of our addressing scheme to aggregate the geographic addresses, as described in Section \ref{sec:prev_work:geo_addressing}, it is possible to advertise an entire network as a single coverage area. This enables geographic routing on a large scale, as each network would not be represented by a single or even multiple coverage areas per router, but by a single unified area.

As an example we take a network covering an entire city. This network could aggregate the coverage area of the routers in a single address. While this single address might not be completely identical to the coverage area of the individual routers (it will be slightly larger in reality), it allows the network to advertise its area in a single advertisement.
The same holds for any autonomous system, we believe our system could achieve Internet-wide scale geocast using this method.

\section{Evaluation}
\label{sec:evaluation}
In this section, we will describe the evaluation of our proposed path based routing algorithm. We will start with briefly mentioning the tools we used to perform the evaluation followed by the method for destination selection. We then describe how we measure the path based algorithm's link usage and describe the method we used to evaluate our algorithms. We will evaluate how optimal the routing tree constructed by the algorithm is. Our main metric will be the number of links used to construct the tree.

\subsection{Tools}
All our evaluations are run over a set of real-world networks taken from the Topologyzoo \cite{topzoo} unless otherwise noted. Using these networks, we hope to more accurately evaluate performance in real-world scenarios as compared to randomly generated ones.

To analyze our algorithms on these networks we use the network library NetworkX \cite{networkx} for the python programming language. We use this tool to load the network graphs.
We use our own code to handle the route distribution and the packet forwarding analyses.

\subsection{Destination distributions}
For our evaluation we define two categories of destination sets: Geographically scoped and randomly distributed destinations. We believe these two sets cover most realistic use cases.

\subsubsection{Geographically scoped destinations}
In most networks that we have evaluated we observe that the geographical distance between two routers and the network distance (number of hop between them) is closely linked. This observation has led us to believe that within a network most geocast traffic will be geographically scoped in its destination router set.

For our geographically scoped destination set we use each node in the network as a source for every possible geographically scoped destination set. The destinations are selected based on their location, for $n$ destinations we select a node in the network and add the $n-1$ geographically closest nodes to the set. Each node is selected once, duplicate sets are filtered out as they would represent the same destination area. The source is never included in the destination set. The result is that each possible geographically scoped (source, destination) combination in the network is evaluated exactly once.

\subsubsection{Randomly distributed destination}
Because in practice it seems unlikely all geocast destination will be geographically clustered in a network we also evaluate randomly selected destinations. We believe such a situation can occur when a network ($A$) serves multiple other networks (for example $B,C,E$). Let's assume networks $B$,$C$ and $E$ all cover our destination area. It is unlikely that the connections of these networks to $A$ are geographically clustered, thus the randomly distributed destination scenario is also important.

As with the geographically scoped destinations we select each node once as the source. For the destinations we select every possible combination of destinations that do not include the source exactly once.

\begin{figure*}[t]%
	\centering
	\subfloat[Full results \label{fig:loop_example_scatter} ]
	{{\includegraphics[width=0.45\linewidth, clip=true]{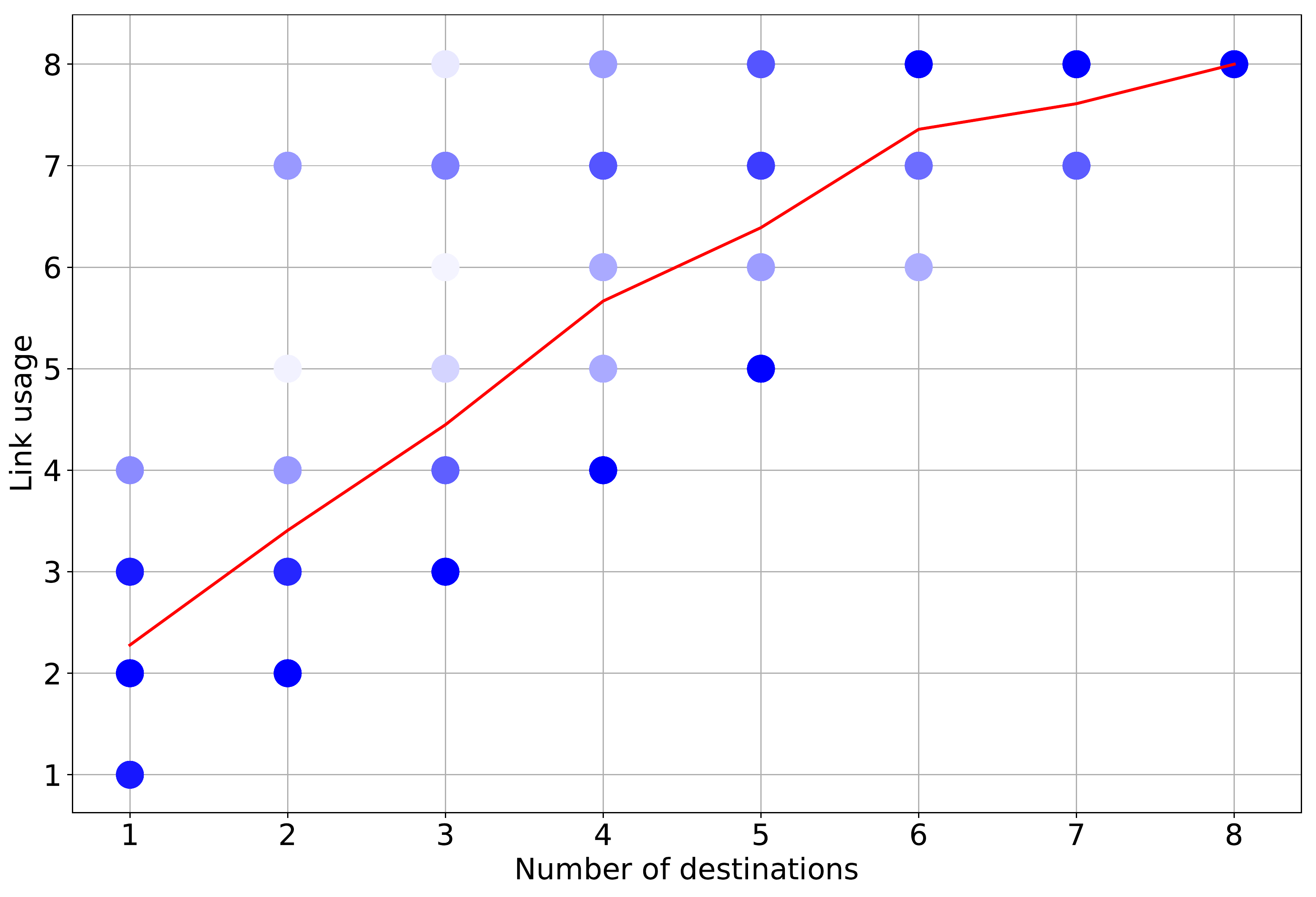} }}%
	\qquad
	\subfloat[Averaged costs \label{fig:loop_example_performance} ]
	{{\includegraphics[width=0.45\linewidth, clip=true]{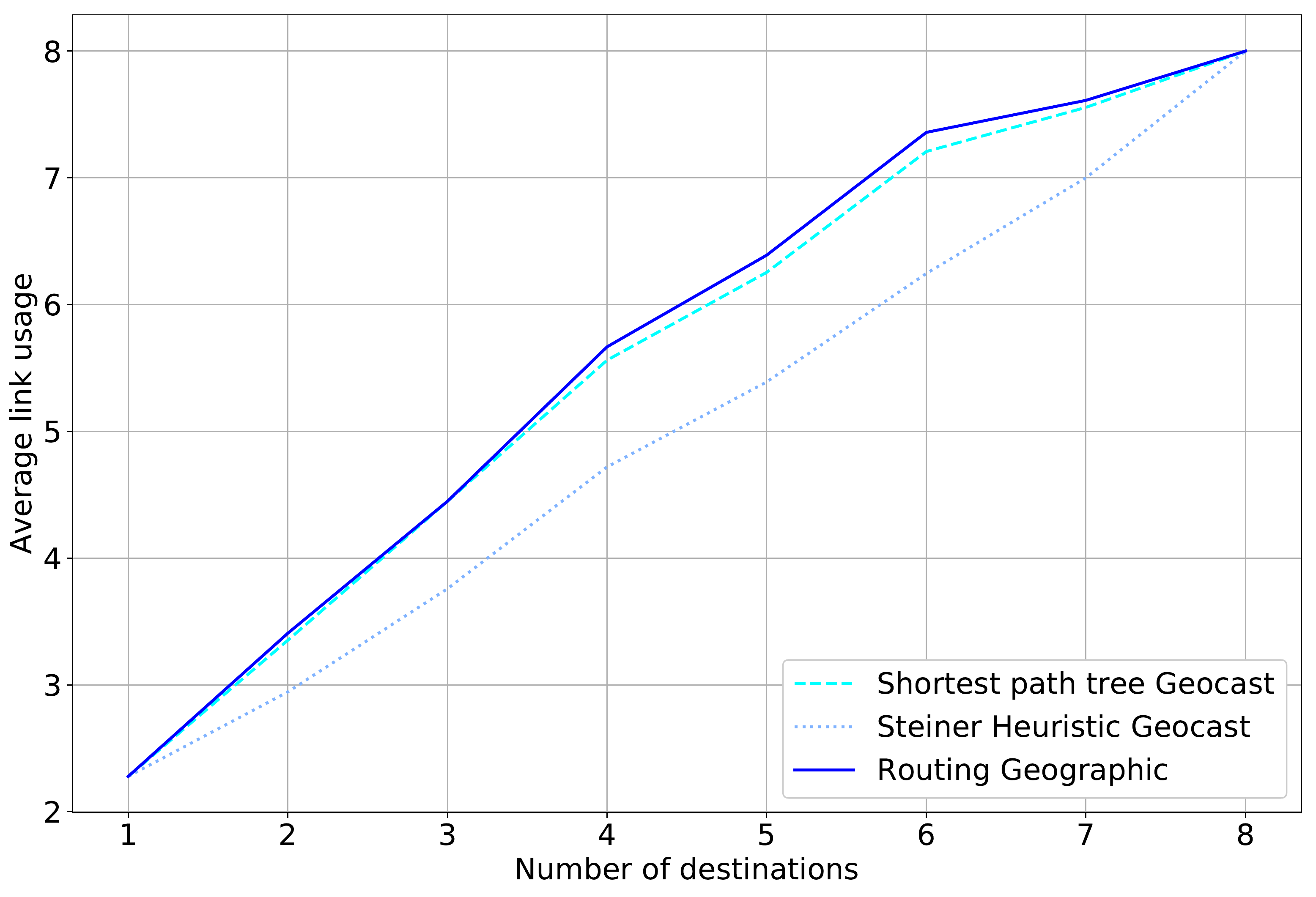} }}%
	\caption{Normalized link cost for the network from Figure \ref{fig:loop_example}}%
	\label{fig:loop_example_plots}%
\end{figure*}

\subsection{Evaluation metrics}
To evaluate the routing performance we look at the resulting link usage in our evaluation graphs. This metric will give us an indication of how efficient the routing algorithm performs its goal of establishing a shortest path forwarding tree. We compare the average number of links used per number of destinations.

We define link usage as the number of links that are used to forward a message from the source to all destinations. If a link was used twice (i.e. in both directions) this counts as two link uses. For example the forwarding tree shown in Figure \ref{fig:algorithm_1} uses 7 links while the more efficient algorithm in Figure \ref{fig:algorithm_4} only uses 5 links. The lower the overall link usage of an algorithm is, the more optimal we consider the forwarding tree.

In Figure \ref{fig:loop_example_scatter}, we show the routing cost in links used over all geographically scoped destination locations in the example network from Figure \ref{fig:loop_example}. The red line represents the average link usage, with the color intensity for the blue dots showing the relative occurrence of the link usage for a certain number of destinations. The effect of the loop inside a loop (explained before in Section \ref{sec:algorithms:path_distance_vector:sp_trees}) can be clearly seen here with the cost of 3 and 4 destinations mainly around 4 and 7, but not 5 and 6.

Figure \ref{fig:loop_example_performance} plots the performance in link cost of the routing algorithm (solid line) against the performance of a perfect shortest path tree (dashed line) and a Steiner tree heuristics algorithm (dotted line) for the same networks as that in Figure \ref{fig:loop_example_scatter}. The solid line corresponds to the red line in Figure \ref{fig:loop_example_scatter}. We can see that the performance of the routing algorithm in terms of links used is close to that of the shortest path tree.

As different networks have different numbers of routers and links, the results for them are not directly comparable. We normalize the link usage to allow us to make this comparison. The normalization of link usage is done by dividing the link usage with the number of links in the network, resulting in a number between 0 and 1. Values above 1 are possible if there are multiple transmission on the same link.

\subsection{Evaluation Method}
We evaluate our routing algorithm by running it on a collection of real world networks taken from the Topology Zoo project \cite{topzoo}.
We initialize every network by performing the route distribution step until the routing table of each router is stable.
This is done by letting the routers exchange information in steps, in each step all routers transmit their path information to all their neighbors. 

We evaluate each possible (source, destination(s)) combination, generated in the way described earlier in this section, in the network by inserting a packet with the given destination(s) at the source router and forwarding it until no router has any operations left to perform.
Forwarding is performed by the path based algorithm described in Section \ref{sec:algorithms:path_distance_vector:forwarding}. Each packet is forwarded on all the link(s) this algorithm returns based on the source, destinations and information from the previous and candidate next hop routers. Similar to the route distributions the forwarding is also performed in steps, in which each step allows all routers to forward a packet if they have any.

This simulation is run for a subset of destinations that are geographically scoped and for randomly distributed destinations as described before.
We then compare the average amount of links used for a given number of destinations to the amount of links used by a shortest path tree for the same (source, destination) combinations.

\subsection{Evaluation Results}

\begin{figure*}[t]%
	\centering
	\subfloat[Geographically scoped \label{fig:norm_link_cost_real} ]
	{{\includegraphics[width=0.45\linewidth, clip=true]{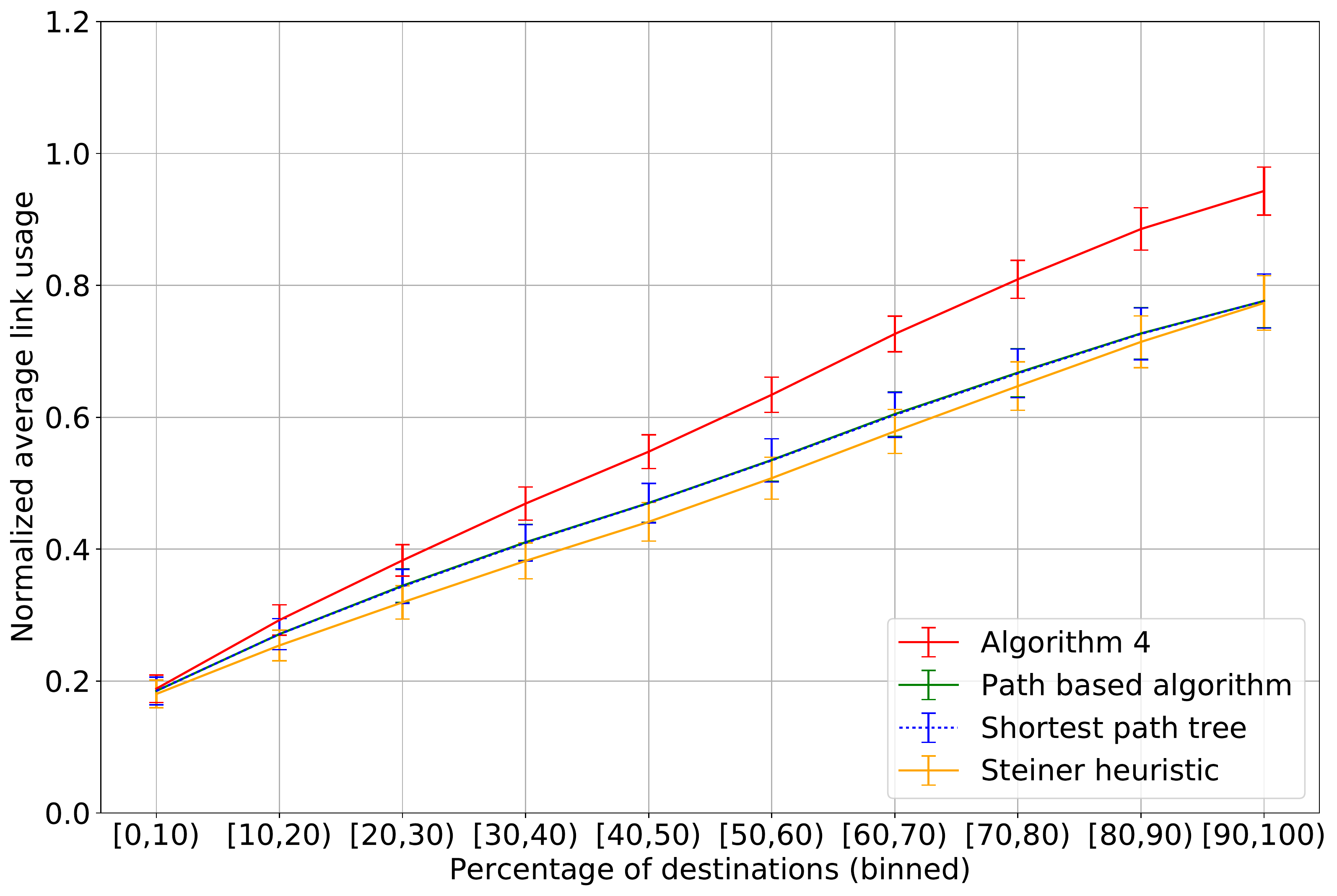} }}%
	\qquad
	\subfloat[Randomly distributed \label{fig:norm_link_cost_multicast} ]
	{{\includegraphics[width=0.45\linewidth, clip=true]{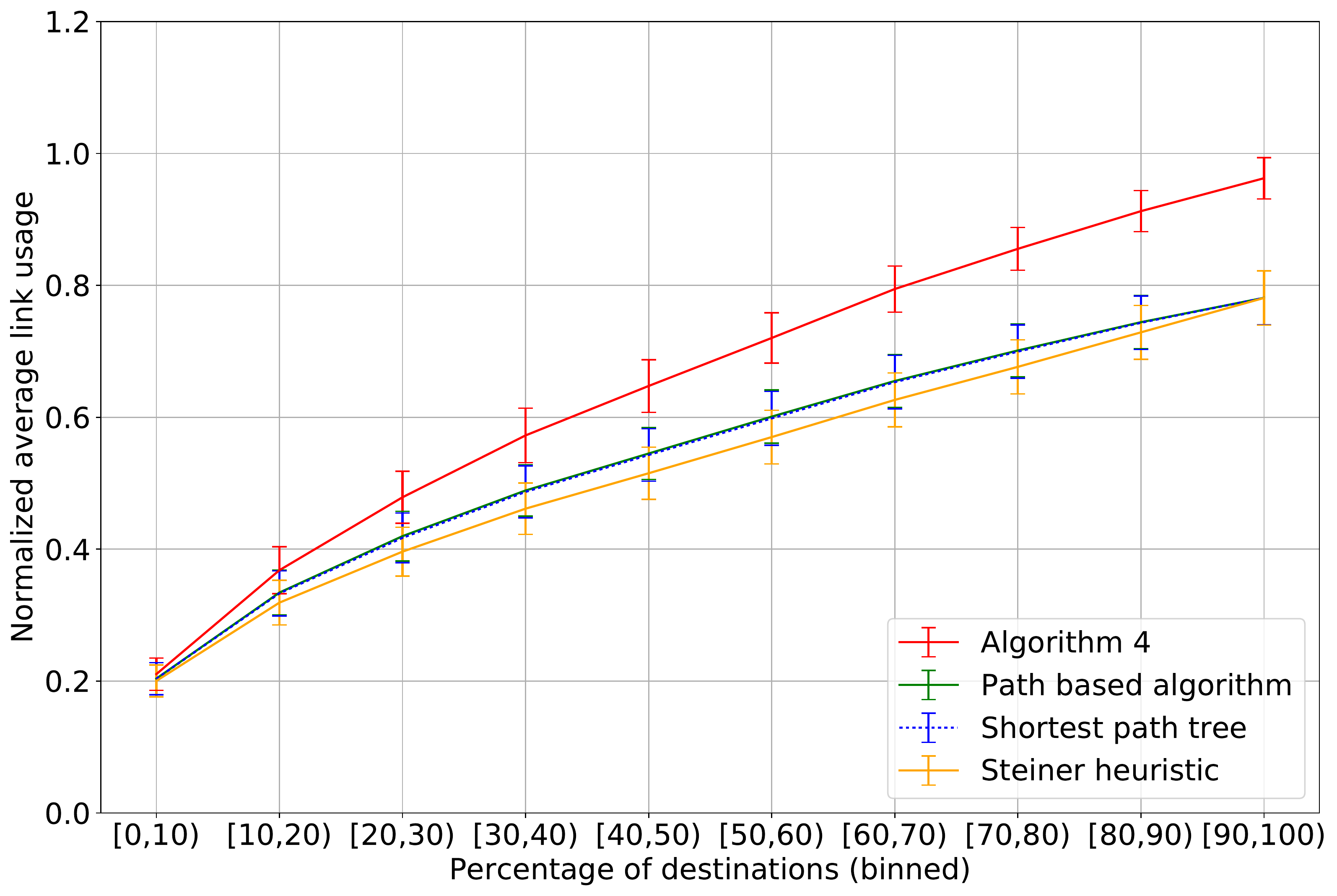} }}%
	\caption{Normalized link cost for real world graphs comparing geocast and multicast}%
	\label{fig:norm_link_cost}%
\end{figure*}

In Figure \ref{fig:norm_link_cost}, we show the normalized link usage on the y-axis. The x-axis represents the normalized number of destinations. This normalization is done by binning the number of destinations for every 10\%. We show algorithm 4, the path based algorithm, shortest path tree and Steiner heuristic normalized over a subset of real world networks. This Figure contains results for a subset of graphs containing the 86 graphs over which we also have a complete set of shortest path and Steiner heuristic results for multicast or randomly distributed locations. This set is limited due to the time needed to evaluate all random combinations in larger networks.

Figure \ref{fig:norm_link_cost_real} shows the geographically scoped results over the 86 graphs while Figure \ref{fig:norm_link_cost_multicast} shows the same values but for randomly distributed destinations. We believe such a scenario could occur in transit networks were the points networks connect to each other do not necessarily correlate with their geographic coverage, especially if these networks cover the same area.

In both cases the line for our path based algorithm and the shortest path tree almost completely overlap. As expected based on our previous work, the average extra link usage compared to the Steiner tree is relatively small. We also note that our best purely DV based algorithm performs reasonably well when the number of destination in a network is small.

\begin{figure*}[t]%
	\centering
	\subfloat[Link usage set against percentage of destinations addressed \label{fig:norm_link_cost_real_all} ]
	{{\includegraphics[width=0.45\linewidth, clip=true]{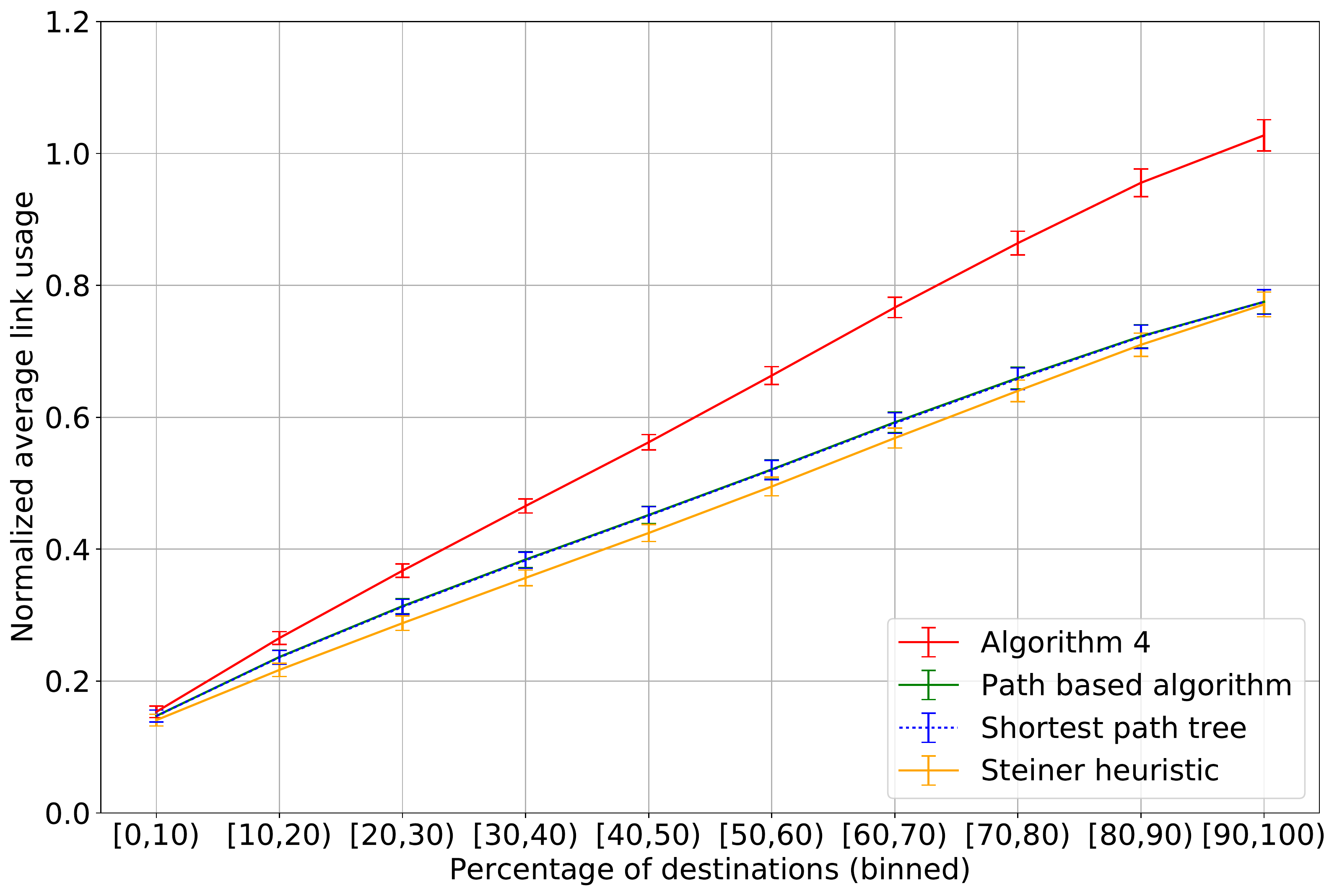} }}%
	\qquad
	\subfloat[Link usage set against the average node degree of the network \label{fig:link_cost_node_deg} ]
	{{\includegraphics[width=0.45\linewidth, clip=true]{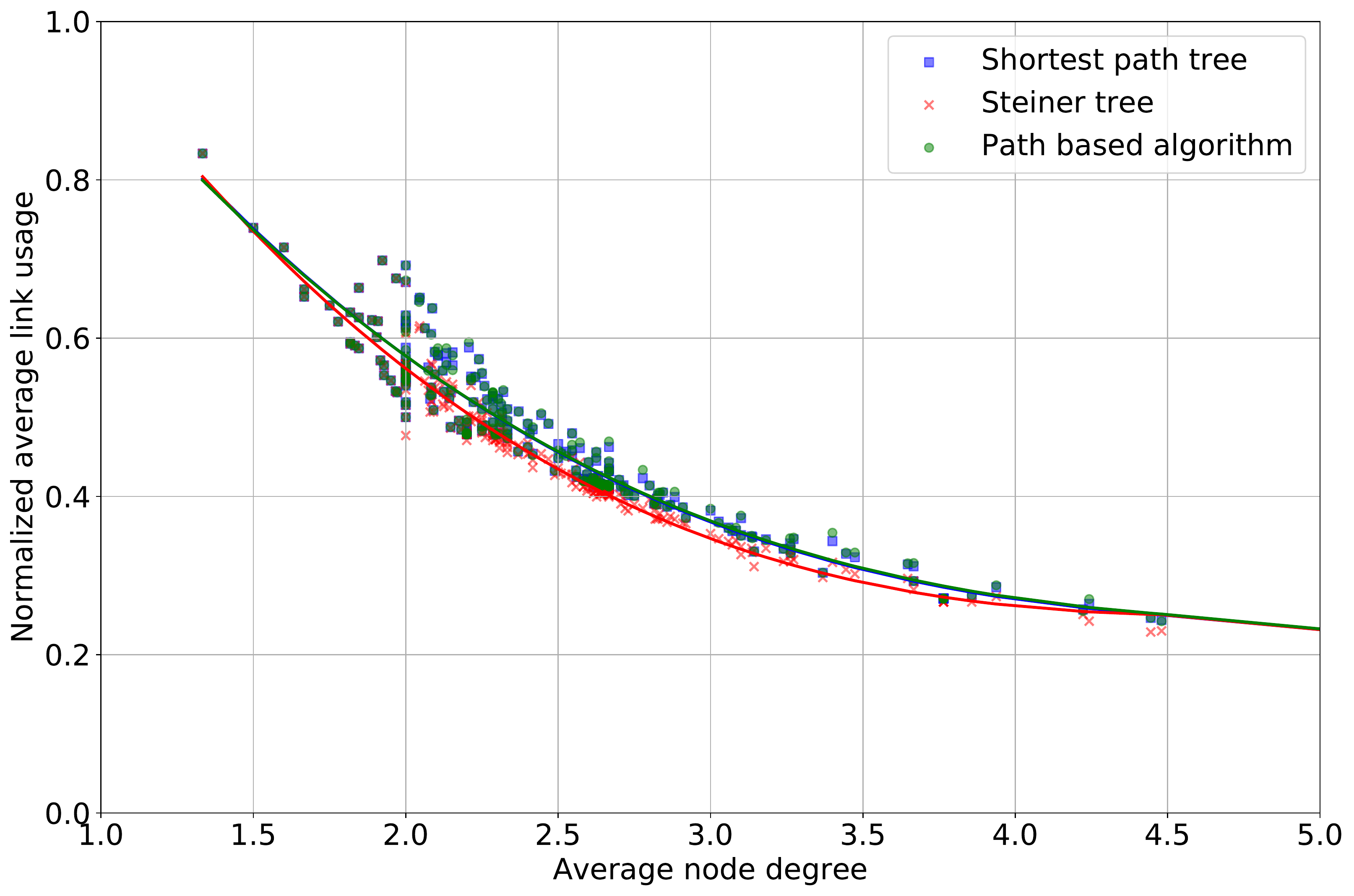} }}%
	\caption{Geographically scoped normalized link usage for all real world graphs}%
	\label{fig:norm_link_cost_full}%
\end{figure*}

Figure \ref{fig:norm_link_cost_full} contains geographically scoped results over the complete set of 227 real-world network graphs we used for the evaluation. Figure \ref{fig:norm_link_cost_real_all} shows results over this set using the same method used for Figure \ref{fig:norm_link_cost_real}. We can see that over this larger set of graphs that also contains larger networks the average cost for our shortest path based algorithms is similar, but the link usage of algorithm 4 is slightly larger implying that its performance might degrade depending on the network size.

In Figure \ref{fig:link_cost_node_deg} we show the normalized average link cost of an entire graph (the average number of links used over all geographic destination combinations divided by the total number of links in a graph) plotted against the average node degree of the network. The average node degree is the average number of links a router has, this is an indication of how well connected a network is. We can see that the average link cost of the routing algorithm is close or equal to that of the shortest path tree. In general, the cost for geographically scoped destinations is close to that of the ideal Steiner tree. We also observe that the more well connected a network is, the lower the average cost to reach a certain destination area.

\subsection{Path knowledge vs. hop knowledge}
Our distance vector based algorithm described in Section \ref{sec:algorithm4} has worse forwarding performance than the algorithm using path knowledge described later. It does however have some benefits over the better performing algorithm:
\begin{itemize}
	\item Lower communications overhead due to DV like cost exchange
	\item Lower lookup complexity
\end{itemize}
The communication overhead depends on the size of the network, the larger the network is, the longer that paths are that our path based algorithm has to communicate through the network. The lookup complexity also depends on the path length, but even with short paths the extra steps required to combine them in a next hop and previous hop view would mean higher complexity than the distance vector based approach

Over the 227 real networks we have evaluated the algorithm in, on average the distance vector based algorithm has 28\% worse performance compared to the more complex algorithm with a standard deviation of 0.26. The best case was identical performance with the worst case 112\% extra transmission.
We can conclude that in some networks the extra transmission overhead could be an acceptable trade-off for the lower computational burden put on the routers themselves. There is no single perfect choice here, the algorithm will have to be selected based on the network. We do observe that the overhead of the distance vector algorithm is lower in smaller networks, and is almost always high in larger networks of more than 15 nodes.

\section{Conclusion}
\label{sec:conclusion}
In this paper, we have presented a purely distance vector and a path based algorithm for geographic routing. We have also evaluated the link usage of these algorithms on a set of real world networks.

Our best distance vector based algorithm performs relatively well, and in the worst case has only 32\% more link cost compared to the shortest path tree. In a situation where the entire network is not addressed this overhead is even lower. 

We have shown that our path based algorithm can construct forwarding trees to multiple destinations that are close in link cost to the shortest path tree from the source to the destinations. Our proposed algorithm establishes forwarding trees that are almost equal to the shortest path tree and close to the optimal Steiner tree in link cost. The algorithm can improve on the distance vector based algorithm, especially in situations where a large number of routers in the network ($>25\%$) are part of the addressed area.

We believe that the distance vector algorithm might actually be preferable in certain situations where the extra computational overhead in the router does not outweigh the extra transmission overhead in the network. We expect that this routing approach combined with a hierarchical approach in which autonomous systems advertise one or more areas will eventually allow Internet-wide geocast to become a reality.

\subsection{Future work}
To achieve Internet-wide geocast, further work will need to be done. Areas of special interest are hierarchical routing, security and last hop distribution methods.

Hierarchical routing is needed so different autonomous systems can distribute coverage and reachability information. Further work is needed to research methods to extend our current routing proposal to this level. 

The are several security issues mostly relating to the possibility of denial of service attacks to the current proposal that need to be addressed. Further work is needed on limiting geocast capabilities to certain users, or limiting the area addressed to prevent wide scale denial of service attacks using geocast.

Work needs to be done for the final hop towards mobile devices, including vehicles. The hop towards end-user devices presents a challenge for geographically scoped communication. There are several different situations in which the method of distribution would need to be tailored to the specific situation. While our addressing method can address small areas, addressing a specific road for example, would still require some form of translation to limit the message to just that road, and not the surrounding area.

We are currently in the process of developing an implementation of our algorithm that can be tested in a virtual environment such as mininet and on actual routing hardware. This will also allow us to evaluate the routing performance in situations where the network is unstable and routes are still converging.

\IEEEtriggeratref{8}
\bibliographystyle{IEEEtran}
\bibliography{main}

\end{document}